\documentstyle[12pt,epsf,feynmp,epsfig]{article}
\def\pls{\makebox(0,0){$+$}}

\def\ssp{\makebox(0,0)
    {\thinlines\put(-.1,0){\line(1,0){.2}}\put(0,-.1){\line(0,0){.2}}}}
\def\ssm{\makebox(0,0){\put(-.1,0){\thinlines\line(1,0){.2}}}}

\def\propagatorR#1#2#3#4{\begin{picture}(1.4,.8)#3
    \if#4r
    \put(1.3,.1){\circle{.2}}\put(1.1,.3){$R$}\else
    \put(.1,.1){\circle{.2}}\put(-0.1,.3){$R$}\fi
    \put(1.3,0.1){\vector(-1,0){.8}}\put(.1,0.1){\line(1,0){.6}}
    \put(.2,-.15){\makebox(0,0){#1}}
    \put(1.2,-.15){\makebox(0,0){#2}}\end{picture}}
\def\propagatorA#1#2#3#4{\begin{picture}(1.4,.6)#3
    \if#4r
    \put(1.3,.1){\circle{.2}}\put(1.1,.3){$R$}\else
    \put(.1,.1){\circle{.2}}\put(-0.1,.3){$R$}\fi
    \put(0.1,0.1){\vector(1,0){.8}}\put(.7,0.1){\line(1,0){.6}}
    \put(.2,-.15){\makebox(0,0){#1}}
    \put(1.2,-.15){\makebox(0,0){#2}}\end{picture}}
\def\self#1#2#3{\begin{picture}(2.2,.8)\thicklines
    \put(2.3,0.1){\vector(-1,0){1.3}}\put(-.1,0.1){\line(1,0){1.3}}
    \put(1.1,0.1){\oval(2,1)}
    \put(0,-0.15){\makebox(0,0){#1}}\put(2.2,-0.15){\makebox(0,0){#2}}
    \if#3R {\put(0.1,0.1){\circle{.2}}}\else
    \if#3A {\put(2.1,0.1){\circle{.2}}}\else
    {\put(1.1,0.1){\makebox(0,0){#3}}}
    \fi\fi\end{picture}}

\linethickness{1.0pt}

\def\gsand4{\begin{picture}(38,0)
\put(2,2){
     \thicklines
     \put(2,0){\circle*{2.00}}
     \put(32,0){\circle*{2.00}}
     \bezier{216}(2,0)(17,-10)(32,0)
     \bezier{216}(2,0)(17,+10)(32,0)
     \bezier{284}(2,0)(17,-25)(32,0)
     \bezier{284}(2,0)(17,+25)(32,0)}
     \end{picture}}
\def\ssand3{\begin{picture}(38,0)
\put(2,2){
     \thicklines
     \put(2,0){\line(1,0){30}}
     \put(2,0){\circle*{4.00}}
     \put(32,0){\circle*{4.00}}
     \bezier{284}(2,0)(17,-25)(32,0)
     \bezier{284}(2,0)(17,+25)(32,0)}
     \end{picture}}
\def\sphisand2phi{\begin{picture}(42,0)
\put(2,2){
     \thicklines
     \put(2,0){
     \put(2,0){\line(1,0){8}}
     \put(24,0){\line(1,0){8}}
     \put(10,0){\makebox(0,0){$\bigotimes$}}
     \put(24,0){\makebox(0,0){$\bigotimes$}}
     \put(2,0){\circle*{4.00}}
     \put(32,0){\circle*{4.00}}
     \bezier{284}(2,0)(16,-25)(32,0)
     \bezier{284}(2,0)(16,+25)(32,0)}}
     \end{picture}}
\def\gphisand3phi{\begin{picture}(53,0)
\put(2,2){
     \thicklines
     \put(3,0){\line(1,0){44}}
     \put(3,0){\makebox(0,0){$\bigotimes$}}
     \put(47,0){\makebox(0,0){$\bigotimes$}}
     \put(8,0){
     \put(2,0){\circle*{2.00}}
     \put(32,0){\circle*{2.00}}
     \bezier{284}(2,0)(16,-25)(32,0)
     \bezier{284}(2,0)(16,+25)(32,0)}}
     \end{picture}}

\def\DSa{\begin{picture}(0,0)\thicklines
\put(0,0){\oval(1.5,1)}
\put(0,0){\makebox(0,0){$-\ii\Sa$}}\end{picture}}
\def\GlnG0Sa{
\begin{picture}(4.3,1.5)
\put(0.5,.1){
\put(.5,0){\DSa}\put(3,0){\DSa}\put(1.75,1){\DSa}
\put(1.75,-1){\makebox(0,0){\dots\dots}}
\put(1,.5){\oval(1,1)[lt]}\put(2.5,.5){\oval(1,1)[rt]}
\put(1,-.5){\oval(1,1)[lb]}\put(2.5,-.5){\oval(1,1)[rb]}}
\end{picture}}
\def\GGaSa{
\begin{picture}(2.5,2.)
\thicklines\put(0.5,.1){
\put(.5,0){\DSa}\put(2,-.5){\line(0,1){1}}
\put(1.,1){\line(1,0){0.5}}\put(1.,-1){\line(1,0){0.5}}
\put(1,.5){\oval(1,1)[lt]}\put(1.5,.5){\oval(1,1)[rt]}
\put(1,-.5){\oval(1,1)[lb]}\put(1.5,-.5){\oval(1,1)[rb]}}
\end{picture}}
\def\vhight#1{\vphantom{\left(\begin{picture}(0,#1)\end{picture}\right)}
}
\def\Dclosed#1#2{
\begin{picture}(2,.8)\put(0,.1){#2
\put(1,0){\circle{1.414}}
\put(1,-.707){\line(0,1){1.414}}\put(.293,0){\line(1,0){1.414}}
\put(0.5,-0.5){\line(0,1){1}}\put(1.5,-0.5){\line(0,1){1}}
\put(0.5,-0.5){\line(1,0){1}}\put(0.5,0.5){\line(1,0){1}}}
\put(1.5,-.8){$#1$}
\end{picture}}
\def\loopb#1{
\unitlength 1.00mm
\thicklines
\begin{picture}(17.50,13.00)
\put(2.00,1.00){\circle*{1.00}}
\put(17.00,1.00){\circle*{1.00}}
\bezier{68}(2.00,1.00)(9.50,5.00)(17.00,1.00)
\bezier{68}(2.00,1.00)(9.50,-3.00)(17.00,1.00)
\bezier{112}(2.00,1.00)(9.50,13.00)(17.00,1.00)
\bezier{112}(2.00,1.00)(9.50,-11.00)(17.00,1.00)
\put(9.93,-1.07){\vector(1,0){0.2}}
\put(9.00,-1.07){\line(1,0){0.93}}
\put(9.93,2.93){\vector(1,0){0.2}}
\put(8.93,2.93){\line(1,0){1.00}}
\put(9.13,6.93){\vector(-1,0){0.2}}
\put(10.07,6.93){\line(-1,0){0.93}}
\put(9.20,-5.00){\vector(-1,0){0.2}}
\put(10.07,-5.07){\line(-1,0){0.87}}
\if#1c \multiput(9.5,-8.)(0,6.5){3}{\line(0,1){5}}\fi
\end{picture}}
\def\loopc#1{ 
\unitlength 1.00mm
\thicklines
\begin{picture}(18.54,9.47)
\put(2.00,-5.00){\circle*{0.93}}
\put(18.00,-5.00){\circle*{1.07}}
\put(10.00,9.00){\circle*{0.94}}
\bezier{72}(2.00,-5.00)(10.00,-1.00)(18.00,-5.00)
\bezier{72}(2.00,-5.00)(10.00,-9.00)(18.00,-5.00)
\bezier{80}(2.00,-5.00)(2.00,6.00)(10.00,9.00)
\bezier{72}(2.00,-5.00)(9.07,-0.80)(10.00,9.00)
\bezier{72}(18.00,-5.00)(11.07,-0.93)(10.00,9.00)
\bezier{76}(18.00,-5.00)(18.13,5.07)(10.00,9.00)
\put(9.93,-7.00){\vector(-1,0){0.2}}
\put(11.00,-7.00){\line(-1,0){1.07}}
\put(3.80,3.27){\vector(1,2){0.2}}
\multiput(3.40,2.47)(0.10,0.20){4}{\line(0,1){0.20}}
\put(16.73,2.33){\vector(2,-3){0.2}}
\multiput(16.20,3.13)(0.11,-0.16){5}{\line(0,-1){0.16}}
\put(10.07,-3.00){\vector(1,0){0.2}}
\put(9.07,-3.00){\line(1,0){1.00}}
\put(12.53,0.60){\vector(-2,3){0.2}}
\multiput(13.00,-0.20)(-0.12,0.20){4}{\line(0,1){0.20}}
\put(7.60,0.67){\vector(-2,-3){0.2}}
\multiput(8.00,1.33)(-0.10,-0.17){4}{\line(0,-1){0.17}}
\if#1c \multiput(8.5,-9.5)(3.5,5.25){3}{\thinlines\line(2,3){2.5}}\fi
\end{picture}}
\def\ssp{\makebox(0,0)
  {\thicklines\put(-.1,0){\line(1,0){.2}}\put(0,-.1){\line(0,0){.2}}}}
\def\ssm{\makebox(0,0){\put(-.1,0){\thicklines\line(1,0){.2}}}}

\def\cal{\mathcal}
\def\pls{\makebox(0,0){$+$}}

\def\ssp{\makebox(0,0)
    {\thinlines\put(-.1,0){\line(1,0){.2}}\put(0,-.1){\line(0,0){.2}}}}
\def\ssm{\makebox(0,0){\put(-.1,0){\thinlines\line(1,0){.2}}}}
\def\loopb#1{
\unitlength 1.00mm
\thicklines
\begin{picture}(17.50,13.00)
\put(2.00,1.00){\circle*{1.00}}
\put(17.00,1.00){\circle*{1.00}}
\bezier{68}(2.00,1.00)(9.50,5.00)(17.00,1.00)
\bezier{68}(2.00,1.00)(9.50,-3.00)(17.00,1.00)
\bezier{112}(2.00,1.00)(9.50,13.00)(17.00,1.00)
\bezier{112}(2.00,1.00)(9.50,-11.00)(17.00,1.00)
\put(9.93,-1.07){\vector(1,0){0.2}}
\put(9.00,-1.07){\line(1,0){0.93}}
\put(9.93,2.93){\vector(1,0){0.2}}
\put(8.93,2.93){\line(1,0){1.00}}
\put(9.13,6.93){\vector(-1,0){0.2}}
\put(10.07,6.93){\line(-1,0){0.93}}
\put(9.20,-5.00){\vector(-1,0){0.2}}
\put(10.07,-5.07){\line(-1,0){0.87}}
\if#1c \multiput(9.5,-8.)(0,6.5){3}{\line(0,1){5}}\fi
\end{picture}}
\def\loopc#1{ 
\unitlength 1.00mm
\thicklines
\begin{picture}(18.54,9.47)
\put(2.00,-5.00){\circle*{0.93}}
\put(18.00,-5.00){\circle*{1.07}}
\put(10.00,9.00){\circle*{0.94}}
\bezier{72}(2.00,-5.00)(10.00,-1.00)(18.00,-5.00)
\bezier{72}(2.00,-5.00)(10.00,-9.00)(18.00,-5.00)
\bezier{80}(2.00,-5.00)(2.00,6.00)(10.00,9.00)
\bezier{72}(2.00,-5.00)(9.07,-0.80)(10.00,9.00)
\bezier{72}(18.00,-5.00)(11.07,-0.93)(10.00,9.00)
\bezier{76}(18.00,-5.00)(18.13,5.07)(10.00,9.00)
\put(9.93,-7.00){\vector(-1,0){0.2}}
\put(11.00,-7.00){\line(-1,0){1.07}}
\put(3.80,3.27){\vector(1,2){0.2}}
\multiput(3.40,2.47)(0.10,0.20){4}{\line(0,1){0.20}}
\put(16.73,2.33){\vector(2,-3){0.2}}
\multiput(16.20,3.13)(0.11,-0.16){5}{\line(0,-1){0.16}}
\put(10.07,-3.00){\vector(1,0){0.2}}
\put(9.07,-3.00){\line(1,0){1.00}}
\put(12.53,0.60){\vector(-2,3){0.2}}
\multiput(13.00,-0.20)(-0.12,0.20){4}{\line(0,1){0.20}}
\put(7.60,0.67){\vector(-2,-3){0.2}}
\multiput(8.00,1.33)(-0.10,-0.17){4}{\line(0,-1){0.17}}
\if#1c \multiput(8.5,-9.5)(3.5,5.25){3}{\thinlines\line(2,3){2.5}}\fi
\end{picture}}
\def\ssp{\makebox(0,0)
  {\thicklines\put(-.1,0){\line(1,0){.2}}\put(0,-.1){\line(0,0){.2}}}}
\def\ssm{\makebox(0,0){\put(-.1,0){\thicklines\line(1,0){.2}}}}

\newcommand{\di}{{\mathrm d}}
\newcommand{\Tr}{{\mathrm{Tr}}}
\newcommand{\ii}{{\mathrm i}}

\renewcommand{\and}{\quad{\mathrm{and}}\quad}

\renewcommand{\Re}{{\mathrm{Re}}}
\renewcommand{\oint}{\int_{\cal C}}
\renewcommand{\Im}{{\mathrm{Im}}}

\def\scr#1{\mbox{\scriptsize #1}}

\def\vec#1{\mbox{\boldmath $#1$}}
\newcommand{\dpi}[1]{\frac{\di^4 #1}{(2\pi)^4}}                
\newcommand{\Pbr}[1]{\left\{#1\right\}}                    
\newlength{\charwidth}

\newcommand{\lap}%
{\raisebox{-0.5ex}{$\stackrel{\scriptstyle <}{\scriptstyle \sim}$}}
\newcommand{\gap}%
{\raisebox{-0.5ex}{$\stackrel{\scriptstyle >}{\scriptstyle \sim}$}}

\def\Gr{G}\def\Se{\Sigma}

\def\Ga{\Gr}
\def\Sa{\Se}

\def\Lg{{\cal L}}
\def\Lgh{\makebox[3.5mm]{${\widehat{\makebox[2mm]{$\Lg$}}}$}\vphantom{L}}
\def\Lint{\Lgh^{\mbox{\scriptsize int}}}

\def\A{A}
\def\Gm{\Gamma}
\def\F{F}                             
\def\Ft{\widetilde{F}}                 
\def\Fd{F}                             
\def\Fdt{\widetilde{F}}                
\def\fd{f}                             

\def\tp{\widetilde{p}}\def\tm{\widetilde{m}}\def\tW{\widetilde{W}} 

\def\Ge{\Gamma_{\scr{in}}}     

\def\Ld{\Gamma_{\scr{out}}}
\def\Ldt{\Gamma_{\scr{in}}}

\def\Do{{\cal D}}

\def\interaction{\makebox(0,0){\put(0,0){\interact}
    \put(0,.95){\ssp}\put(0,-.95){\ssm}
    \put(0,.125){\ssp}\put(0,-.125){\ssm}}}
\def\interact{\makebox(0,0){\put(0,.5){\fullbox}
    \thicklines\put(0,0){\oval(.75,.5)}
    \put(0,-.5){\fullbox}} }
\def\til2loop{\put(0,0){\oneloopvertex}\put(4.,0){\pls}
      \put(4.75,0){\oneloopvertex}\put(6.375,0){\interaction}
      \put(9,0){\pls}}

\def\photon{\thinlines\multiput(0,0)(.2,0){3}{\line(1,0){0.1}}}
\def\oneloopvertex{
    \put(1.625,0){\thicklines\oval(2.0,1.5)}
    \put(0,0){\photon}\put(0.35,0.3){\ssp} 
    \put(0.625,0){\circle*{.25}}\put(2.625,0){\circle*{.25}}
    \put(2.75,0){\photon}\put(2.9,0.3){\ssm}}
\def\fullbox{\makebox(0,0){\rule{1.5mm}{3mm}}}

\def\Blue{}
\def\Red{}\def\Green{}
\def\citerange[#1-#2]{[\citem[#1,@]-\citem[#2,@]]}
\def\citem[#1,#2]{\csname b@#1\endcsname\if @#2{}\else ,\citem[#2]\fi}
\begin{document}
{\large 
{\title{Soft Modes, Resonances and Quantum Transport\footnote{submitted to
Phys. At. Nucl. (Rus.), the volume dedicated to the memory of A.B. Migdal} }
\author{
Yu. B. Ivanov\footnote{permanent address:
\it Kurchatov Institute, Kurchatov sq. 1, Moscow 123182,
Russia},
J. Knoll,
H. van Hees,
D. N. Voskresensky\footnote{permanent address:
\it Moscow Institute for Physics and Engineering, 
Kashirskoe sh. 31, Moscow 115409, Russia}
}
\maketitle
}
{\normalsize 
\begin{center}
{\it {Gesellschaft f\"ur Schwerionenforschung mbH, Planckstr. 1,
64291 Darmstadt, Germany}}
\end{center}
{
\begin{abstract}
{ Effects of the propagation of particles, which have a
  finite life-time and an according width in their mass spectrum, 
  are discussed
  in the context of transport description. First, the
  importance of coherence effects (Landau--Pomeranchuk--Migdal effect)
  on production and absorption of field quanta in non-equilibrium dense matter
  is considered. It is shown that classical diffusion and Langevin results
  correspond to re-summation of certain field-theory diagrams formulated in
  terms of full non-equilibrium Green's functions. Then the   general
  properties of broad resonances in dense and hot systems are
  discussed in the framework of a self-consistent and conserving 
  $\Phi$-derivable method of Baym   at the examples of the $\rho$-meson  
  in hadronic matter and the pion in dilute nuclear matter.   Further we 
  address the problem of a transport description that properly accounts for
  the damping width of the particles. 
  The $\Phi$-derivable method generalized to the real-time contour 
  provides a self-consistent and conserving kinetic scheme. 
  We derive a 
  generalized expression for the non-equilibrium kinetic entropy flow, which
  includes corrections from fluctuations and mass-width effects. In special
  cases an $H$-theorem is proved. Memory effects in collision terms give 
  contributions to the kinetic entropy flow that in the Fermi-liquid case
  recover the famous bosonic type $T^3 \ln T$ correction 
  to the specific heat of liquid Helium-3. 
  At the example of the pion-condensate phase transition 
  in dense nuclear matter we demonstrate important part played by the width
  effects  within the quantum transport.  }
\end{abstract}
\begin{fmffile}{migdal-d}
\fmfset{thin}{1.3pt}
\fmfset{arrow_len}{3mm}

Quasiparticle representations in many-body theory
were designed by Landau, Migdal, Galitsky and others, see refs 
~\cite{Landau,Mqp,M,LP1981}. This concept was first elaborated at the example
of low-lying particle-hole excitations in Fermi liquids.
A.B. Migdal was the first to apply these methods to description of various
nuclear phenomena and construction of a closed semi-microscopic approach that
is now usually called "Theory of finite Fermi systems"~\cite{M}. 
The need for explicit treatment of soft
modes  within this approach stimulated him to generalize this concept to 
include  pion and $\Delta$ excitations. A.B. Migdal suggested variety of
interesting effects like softening of the pion mode in nuclei,
pion condensation in dense nuclear and neutron star
matter and  possible existence of abnormal nuclei glued by pion condensate
~\cite{Migdal1971,Migdal1972,Migdal1973,Migdl}. 
These ideas stimulated further development of pion 
physics with applications to many phenomena in atomic nuclei, neutron stars
and heavy-ion collisions, see \cite{Migdl,MSTV,Vos93} and refs therein. 
In this paper we would like to briefly review recent developments of some of
the above ideas as applied to heavy-ion physics. 

With the aim to describe the collision of two nuclei at intermediate and
high energies one is confronted with the fact that the dynamics has to include
particles like the in-medium excitation with the pion quantum numbers, as well
as the delta and rho-meson resonances with life-times of less than
2 fm/c or equivalently with damping widths above 100 MeV. Also the collisional
damping rates deduced from presently used transport codes are of the same 
order, whereas typical mean temperatures 
range between 50 to 150 MeV depending on beam energy. Thus, the damping width
of most of the constituents in the system can by no means be treated as a
perturbation.  As a consequence the mass spectrum of the particles in the
dense matter is no longer a sharp delta function but rather acquires a width
due to collisions and decays. One thus comes to a picture which unifies {\em
  resonances}, which have already a width in vacuum due to decay modes, with
the ''states'' of particles in dense matter, which obtain a width due to
collisions (collisional broadening). Landau, Pomeranchuk and Migdal were
first to demonstrate importance of finite-width effects in multi-particle
scattering \cite{LandauP,Migdal}. Such a coherence scattering effect,  
known now as Landau--Pomeranchuk--Migdal effect, was recently observed at
Stanford accelerator~\cite{eSLAC}.

The theoretical concepts for a proper many-body description in terms of a real
time non-equilibrium field theory have been devised by
Schwinger~\cite{Schw}, Kadanoff and Baym~\cite{Kad62}, and
Keldysh~\cite{Keld64} in the early sixties. 
However a proper dynamical scheme in terms of a transport concept that deals
with unstable particles, like resonances, is
still lacking. Rather ad-hoc recipes are in use that sometimes violate basic
requirements as given by fundamental symmetries and conservation laws,
detailed balance or thermodynamic consistency. The problem of conserving
approximation has first been addressed by Baym and Kadanoff \cite{KadB,Baym}.
They started from an equilibrium description in the imaginary-time formalism
and discussed the response to external disturbances.  Baym, in particular,
showed \cite{Baym} that any approximation, in order to be conserving, must be
so-called $\Phi$-derivable. It turned out that the $\Phi$ functional required
is precisely the auxiliary functional introduced by Luttinger and Ward
\cite{Luttinger} (see also ref. \cite{Abrikos}) in connection with the
thermodynamic potential. In non-equilibrium case the problem of conserving
approximations is even more severe than in near-equilibrium linear-response
theory \cite{IKV,IKV99}.

In this review we discuss recent developments of the transport theory beyond the
quasiparticle approximation and consequences of the propagation of particles
with short life-times in hadron matter. First we consider few examples of
equilibrium systems which clearly indicate that treatment beyond the
quasiparticle approximation is really needed. We start with a genuine
problem related to the occurrence of broad damping width, i.e. the soft-mode
problem (Landau--Pomeranchuk--Migdal effect). This is the 
direct radiation of quanta from a piece of a dense medium~\cite{KV}.
Classically this problem is formulated as coupling of a
coherent classical field, e.g., the Maxwell field, to a stochastic source
described by the Brownian motion of a charged particle. In this case the
classical current-current correlation function, can be obtained in closed
analytical terms and discussed as a function of the macroscopic transport
properties, the friction and diffusion coefficient of the Brownian particle.
We show that this result corresponds to a partial re-summation of photon
self-energy diagrams in the real-time formulation of field theory. 
Subsequently, properties of particles with broad damping width are
illustrated at the examples of the $\rho$-meson in dense matter at finite
temperature \cite{Hees} and the pion in
the limit of a dilute nuclear matter \cite{Weinhold}. 
The question of consistency becomes especially important for a multi-component
system like $\pi N\Delta\rho,...$, 
where the properties of one species can change due to the presence of
interactions with the others and {\em vice versa}. The ''{\em vice versa}'' is
very important and corresponds to the principle of {\em actio} = {\em
  re-actio}. This implies that the self-energy of one species cannot be
changed through the interaction with other species without affecting the
self-energies of the latter ones also. The $\Phi$-derivable 
method of Baym~\cite{Baym} offers a
natural and consistent way to account for this principle. 

Then we address the question 
how particles with broad mass-width can be described
consistently within a transport picture \cite{IKV,IKV99}.  
We argue that the Kadanoff--Baym equations in the first gradient
approximation together with the $\Phi$-functional method of Baym~\cite{Baym}
provide a proper self-consistent
approach for kinetic description of systems of particles with a
broad mass-width. We argue that after gradient expansion the full set of 
equations describing quantum transport contains two equations, 
the differential generalized kinetic equation for a
distribution function in 8-$(X,p)$-space and the algebraic
equation for the spectral density. Other equations are resolved. 
We discuss the problems of concerning charge
and energy--momentum conservation, thermodynamic consistency, memory effects
in the collision term and the growth of entropy in specific
cases. Finally, we demonstrate finite-width effects in quantum kinetic
description at the example of pion condensation, where the 
width of soft pionic excitations due to their decay into particle-hole
pairs governs the dynamics of the phase transition in the isospin-symmetric
nuclear matter. 

We use the units $\hbar =c =1$.
For simplicity we treat 
fermions non-relativistically whereas  bosons (mesons) are
treated as relativistic
particles.

\section{Bremsstrahlung from Classical Sources}
For a clarification of the soft mode problem following \cite{KV}
we first discuss an example in classical
electrodynamics. We consider a stochastic source, the hard matter, where the
motion of a single charge is described by a diffusion process in terms of a
Fokker--Planck equation for the probability distribution $f$ of
position ${\vec x}$ and velocity ${\vec v}$
\begin{eqnarray}\label{FP}
    \frac{\partial}{\partial t} f({\vec x},{\vec v},t)
    =
    \left({D\Gamma_x^2}\frac{\partial^2}{\partial {\vec v}^2}
    +\Gamma_x \frac{\partial}{\partial {\vec v}}{\vec v}-{\vec
    v}\frac{\partial}{\partial {\vec x}}\right) f({\vec x},{\vec v},t).
\end{eqnarray}
Fluctuations also evolve in time according to this equation, or
equivalently by a random walk process~\cite{KV}, and this way determine
correlations. This
charge is coupled to the Maxwell field. On the assumption of a non-relativistic
source, this case does not suffer from standard pathologies encountered in hard
thermal loop (HTL) problems of QCD, namely the collinear singularities, where
${\vec v}{\vec q}\approx 1$, and from diverging Bose-factors. The advantage of
this Abelian example is that damping can be fully included without violating
current conservation and gauge invariance. This problem
is related to the Landau--Pomeranchuk--Migdal effect of bremsstrahlung in
high-energy scattering~\cite{LandauP,Migdal}.
\begin{figure}
\parbox{7.6cm}{\epsfxsize=7.6cm\epsfbox{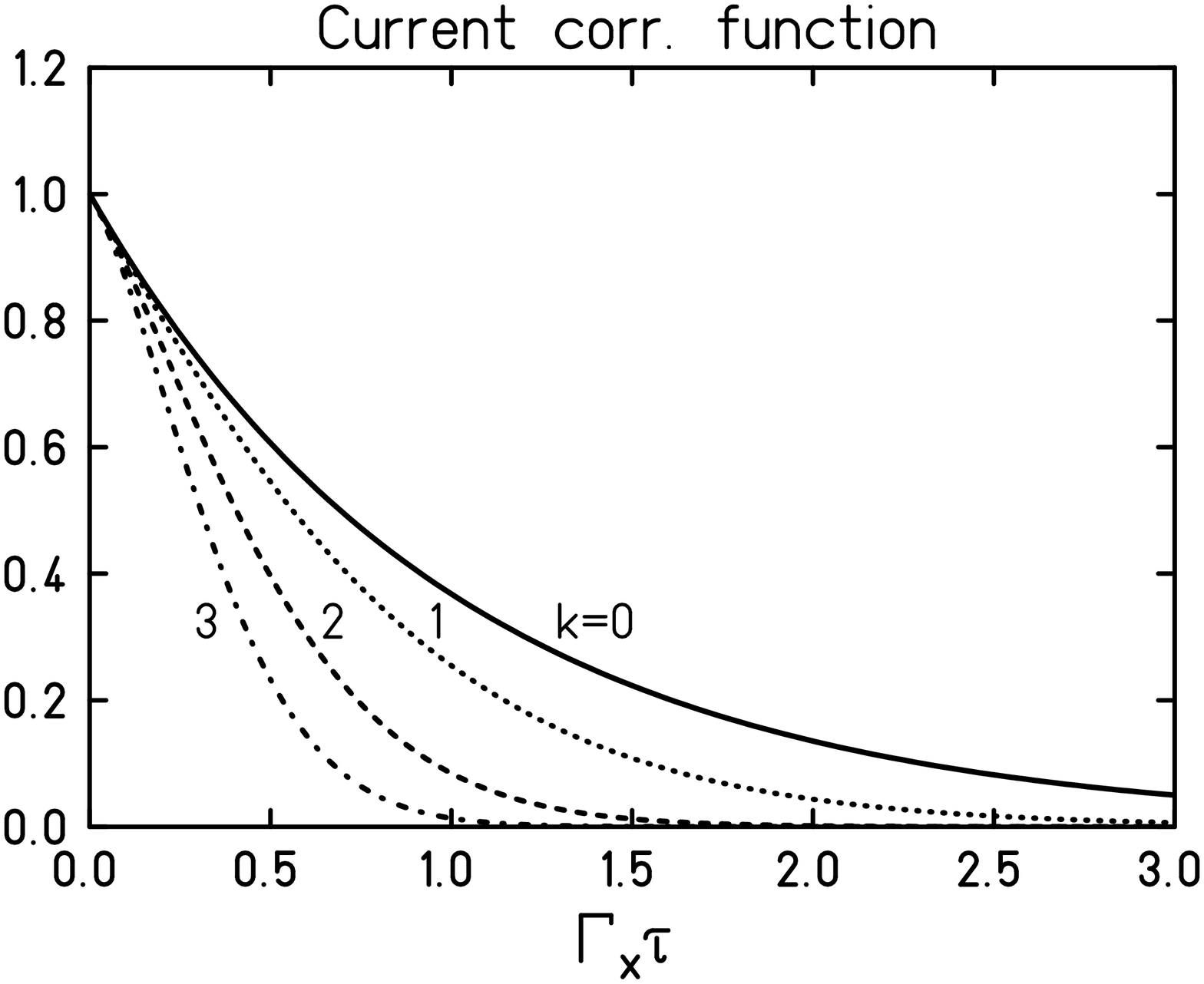}}\hfill
\parbox{7.6cm}{\epsfxsize=7.6cm\epsfbox{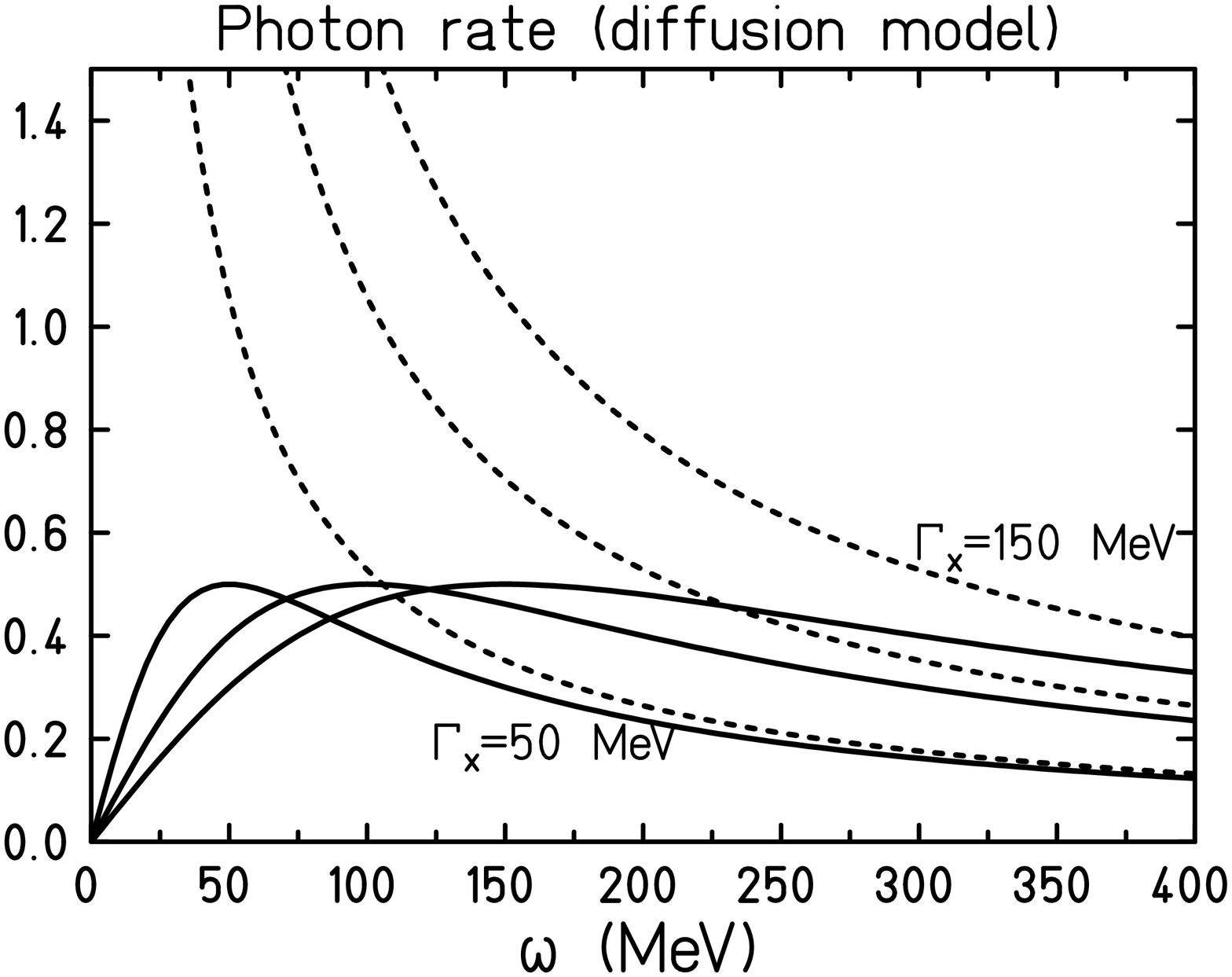}}
\caption{Left: Current-current correlation function
in units of $e^2<v^2>$ as a
function of time (in units of $1/\Gamma_x$) for different values of
the photon momentum $q^2=3k^2\Gamma_x^2/{<v^2>}$ with $k=0,1,2,3$.
Right: Rate of real photons $\di^2
N/(\di\omega\di t)$ in units of $4\pi e^2\left<{\vec v}^2\right>/3$
for a non-relativistic source for $\Gamma_x=$50,100,150 MeV; for
comparison the IQF results (dashed
lines) are also shown.}
\label{CCC}\label{Rate}
\end{figure}

The two macroscopic parameters, the spatial diffusion $D$ and friction 
$\Gamma_x$ coefficients 
determine the relaxation rates of velocities.  In the
equilibrium limit ($t\rightarrow\infty$) the distribution attains a
Maxwell-Boltzmann velocity distribution with the temperature $T=m\left< {\vec
    v}^2\right>/3=mD\Gamma_x$.  The correlation function can be obtained in
closed form and one can discuss the resulting time correlations of the current
at various values of the spatial photon momentum ${\vec q}$, Fig.
\ref{CCC} (details are given in ref.~\cite{KV}).  For the transverse part of
the correlation tensor this correlation decays exponentially as $\sim
e^{-\Gamma_x\tau}$ at ${\vec q}=0$, and its width further decreases with
increasing momentum $q=|{\vec q}|$. The in-medium production rate is given by
the time Fourier transform $\tau\rightarrow\omega$, Fig. \ref{Rate} (right
part). The hard part of the spectrum behaves as intuitively expected, namely,
it is proportional to the microscopic collision rate expressed through 
$\Gamma_x$ (cf. below) and thus can be treated perturbatively by incoherent
quasi-free (IQF) 
scattering prescriptions. However, independently of $\Gamma_x$ the rate
saturates at a value of $\sim 1/2$ in these units around $\omega\sim\Gamma_x$,
and the soft part shows the inverse behavior.  That is, with increasing
collision rate the production rate is more and more suppressed! This is in
line with the picture, where photons cannot resolve the individual
collisions any more. Since the soft part of the spectrum behaves like $
\omega/\Gamma_x$, it shows a genuine non-perturbative feature which cannot be
obtained by any power series in $\Gamma_x$.  For comparison: the dashed lines
show the corresponding IQF yields, which agree with the correct rates for the
hard part while completely fail and diverge towards the soft end of the
spectrum.  For non-relativistic sources $\left<{\vec v}^2\right>\ll 1$ one can
ignore the additional $q$-dependence (dipole approximation; cf. Fig.
\ref{CCC}) and the entire spectrum is determined by one macroscopic scale, the
relaxation rate $\Gamma_x$. This scale provides a quenching factor
\begin{equation}\label{suppression}
C_0(\omega)=\frac{\omega^2}{\omega^2+\Gamma_x^2}\; 
\end{equation}
by which the IQF results have to be corrected in order to account for
the finite collision time effects in dense matter.

The diffusion result represents a re-summation of the microscopic Langevin
multiple collision picture and altogether only macroscopic scales are relevant
for the form of the spectrum and not the details of the microscopic
collisions. Note also that the classical result fulfills the
classical version ($\hbar\rightarrow 0$) of the sum rules discussed in 
refs.~\cite{DGK,KV}.

\section{Radiation at the Quantum Level}
We have seen that at the classical level the problem of radiation from
dense matter can be solved quite naturally and completely at least for
simple examples, and Figs. \ref{CCC} 
display the main
physics. They show, that the {\em damping} of the particles due to
scattering is an important feature, which in particular has to be
included right from the onset.  This does not only assure results
that no longer diverge, but also provides a systematic and convergent
scheme. On the {\em quantum level} such problems require techniques beyond the
standard repertoire of perturbation theory or the quasi-particle
approximation.
In terms of non-equilibrium diagrammatic technique in Keldysh notation, 
the production or absorption rates are given by the self-energy diagrams of 
the type presented in Fig. 2 
\def\ssp{\small $+$}\def\ssm{\small $-$}
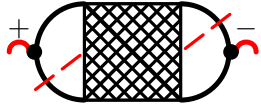
\begin{figure}
\begin{center}
{\unitlength0.8mm
$\;\;\parbox{12.5\unitlength}{
\begin{fmfgraph*}(12.5,16)
\fmfpen{thick}
\fmfstraight
\fmftopn{t}{7}
\fmfbottomn{b}{7}
\fmfleft{l}
\fmfright{r}
\fmf{plain,left}{b7,t7}
\fmfforce{0.3333w,0.5h}{v}
\fmfdot{v}
\fmf{gluon,label=\ssp,l.d=3thick,fore=red}{v,l}
\fmf{plain,width=thin,fore=red}{c1,c2}
\fmf{plain,width=thin,fore=red}{c3,c4}
\fmf{phantom,tension=1.3}{b2,c1}
\fmf{phantom,tension=4}{r,c4}
\fmf{phantom,tension=2}{c2,c3}
\fmfkeep{leftv}
\end{fmfgraph*}}
\parbox{16\unitlength}{
\begin{fmfgraph*}(16,16)
\fmfstraight
\fmftopn{t}{2}
\fmfbottomn{b}{2}
\fmfpoly{hatched}{t2,t1,v1,v2}
\fmf{plain}{t1,t2}
\fmf{plain}{b1,b2}
\end{fmfgraph*}}
\parbox{12.5\unitlength}{
\begin{fmfgraph*}(12.5,16)
\fmfpen{thick}
\fmfstraight
\fmftopn{t}{7}
\fmfbottomn{b}{7}
\fmfright{r}
\fmfleft{l}
\fmf{plain,left}{t1,b1}
\fmfforce{0.6667w,0.5h}{v}
\fmfdot{v}
\fmf{gluon,label=\ssm,fore=red}{r,v}
\fmf{plain,width=thin,fore=red}{c1,c2}
\fmf{plain,width=thin,fore=red}{c3,c4}
\fmf{phantom,tension=1.3}{t6,c1}
\fmf{phantom,tension=4}{l,c4}
\fmf{phantom,tension=2}{c2,c3}
\fmfkeep{rightv}
\end{fmfgraph*}}$}\\[1mm]
\end{center}
\caption{Self-energy diagrams determining the production and absorption rates.
}
\end{figure}
with an in-- and out-going radiating particle 
(e.g. photon) line \cite{VS87,KV}. The hatched loop area
denotes all strong interactions of the source. 
The latter give rise to a
whole series of diagrams.  As mentioned, for the particles of the source, e.g.
the nucleons, one has to re-sum Dyson's equation with the corresponding full
complex self-energy in order to determine the full Green's functions in dense
matter. Once one has these Green's functions together with the interaction
vertices at hand, one could in principle calculate the required diagrams.
However, both the computational effort to calculate a single diagram and the
number of diagrams increase dramatically with the loop order of the
diagrams, such that in practice only lowest-order loop diagrams can be
considered in the quantum case. In certain limits some diagrams drop out.
We could show that in the {\em classical limit},
which in this case implies the hierarchy $\omega,|{\vec q}|,\Gamma\ll T\ll m$
together with low phase-space occupations for the source, i.e.  $f(x,p)\ll 1$,
only the following set of diagrams survives
\begin{equation}\label{classicaldiagram}
\unitlength0.8mm
\parbox{12.5\unitlength}{
\fmfreuse{leftv}}
\parbox{12.5\unitlength}{
\fmfreuse{rightv}}
\;\;+\;\;
\parbox{12.5\unitlength}{
\fmfreuse{leftv}}
\parbox{12.5\unitlength}{
\begin{fmfgraph*}(12.5,16)
\fmfpen{thick}
\fmfstraight
\fmftopn{t}{11}
\fmfbottomn{b}{11}
\fmfright{r}
\fmfleft{l}
\fmfpoly{full}{t7,t5,vt2,vt3}
\fmfpoly{full}{b5,b7,vb3,vb2}
\fmfpoly{full}{vt3,vt2,mt2,mt3}
\fmfpoly{full}{vb2,vb3,mb3,mb2}
\fmf{plain,left}{mb2,mt2}
\fmf{plain,left}{mt3,mb3}
\fmf{plain}{t1,t11}
\fmf{plain}{b1,b11}
\fmfv{label=\ssp,l.d=1.7thick,fore=red}{t6}
\fmfv{label=\ssm,l.d=1.3thick,fore=red}{b6}
\fmf{plain,width=thin,fore=red}{c1,c2}
\fmf{plain,width=thin,fore=red}{c3,c4}
\fmf{phantom,tension=4}{l,c1}
\fmf{phantom,tension=4}{r,c4}
\fmf{phantom,tension=2}{c2,c3}
\fmfkeep{interloop}
\end{fmfgraph*}}
\parbox{12.5\unitlength}{
\fmfreuse{rightv}}
\;\;+\dots+\;\;
\parbox{12.5\unitlength}{
\fmfreuse{leftv}}
\parbox{12.5\unitlength}{
\fmfreuse{interloop}}
\parbox{12.5\unitlength}{
\fmfreuse{interloop}}
\parbox{12\unitlength}{
\begin{fmfgraph*}(12.5,16)
\fmfpen{thick}
\fmfstraight
\fmftopn{t}{11}
\fmfbottomn{b}{11}
\fmfright{r}\fmfleft{l}
\fmf{phantom,label=$\cdots$,label.d=0}{l,r}
\fmf{plain}{t1,t11}
\fmf{plain}{b1,b11}
\end{fmfgraph*}}
\parbox{12.5\unitlength}{
\fmfreuse{interloop}}
\parbox{12.5\unitlength}{
\fmfreuse{rightv}}
\end{equation}
In these diagrams the bold lines denote the full nucleon Green's functions
which also include the damping width, the black blocks represent the effective
nucleon-nucleon interaction in matter, and the full dots, the coupling vertex
to the photon. The thin dashed lines show how the diagrams are to be cut into
two interfering amplitudes. This way each of these diagrams with $n$
interaction loop insertions just corresponds to the $n^{th}$ term in the
corresponding classical Langevin process, where hard scatterings occur at
random with a constant {\em mean collision rate} $\Gamma$. These scatterings
consecutively change the velocity of a point charge from ${\vec v}_0$ to
${\vec v}_1$, to ${\vec v}_2$, $\dots$. In between scatterings the charge
moves freely. For such a multiple collision process the space integrated
current-current correlation function takes a simple Poisson form
\begin{eqnarray}\label{Apoisson}
\ii\Pi^{\mu\nu -+}   &\propto&\int \di^3x_1\di^3x_2
\makebox[3.5cm][l]{$ \left<j^{\nu}({\vec
      x_1},t-\mbox{$\frac{\tau}{2}$})j^{\mu}({\vec 
 x_2},t+\mbox{$\frac{\tau}{2}$})\right>$}\cr
&=& e^2 
   \left<v^{\mu}(0)v^{\nu}(\tau)\right>
=
   e^2e^{-|\Gamma\tau|}\sum_{n=0}^\infty
   \frac{|\Gamma\tau|^n}{n!} \left< v^{\mu}_0 v^{\nu}_n\right>
\end{eqnarray} 
with $v=(1,{\vec v})$. Here $\left<\dots\right>$ denotes the average over
the discrete collision sequence. This form, which one writes down
intuitively, agrees with the analytic result of the quantum correlation
diagrams (\ref{classicaldiagram}) in the limit $f(x,p)\ll 1$ and $\Gamma\ll
T$. 
Fourier
transformed it determines the spectrum in completely regular terms (void of
any infra-red singularities), where each term describes the interference of the
photon being emitted at a certain time or $n$ collisions later.  In special
cases where velocity fluctuations are degraded by a constant fraction $\alpha$
in each collision, such that $\left< {\vec v}_0\cdot{\vec v}_n \right>=
\alpha^n\left< {\vec v}_0\cdot{\vec v}_0 \right>$, one can re-sum the
whole series in Eq. (\ref{Apoisson}) and thus recover the relaxation result with
$2\Gamma_x\left< {\vec v}^2 \right>= \Gamma\left< ({\vec v}_0-{\vec
    v}_1)^2 \right>$ at least for ${\vec q}=0$ and the corresponding
quenching factor (\ref{suppression}). Thus the classical multiple collision
example provides a quite intuitive picture about such diagrams. Further
details can be found in ref.~\cite{KV}.

The above example shows that we have to deal with particle transport
that explicitly takes account of the particle mass-width in order to
properly describe soft radiation from the system.

\section{The $\rho$-meson in Dense Matter}

Following $\Phi$ derivable scheme we will first discuss two examples
within thermo-equilibrium systems.
First we will concern properties of the
$\rho$-meson and their consequences for the $\rho$-decay into
di-leptons \cite{Hees}.  In terms of the non-equilibrium diagrammatic
technique, the exact production rate of di-leptons 
is given by the following formula \unitlength8mm
\unitlength1mm
\begin{eqnarray}\label{dndtdm}
\frac{\di n^{\mbox{e}^+\mbox{e}^-}}{\di t\di m}&=&
\;\;\;\parbox{60mm}{
\begin{fmfgraph*}(60,25)
\fmfpen{thick}
\fmfstraight
\fmftopn{t}{4}
\fmfbottomn{b}{4}
\fmf{fermion,label=$e^+$,tension=0.5}{t1,vl}
\fmf{fermion,label=$e^-$,tension=0.5}{b1,vl}
\fmf{photon,label=$\gamma^*$,l.d=15}{vr,v2}
\fmf{gluon,label=$\rho$,l.d=15}{v2,v1}
\fmf{photon,label=$\gamma^*$,l.d=15}{v1,vl}
\fmf{fermion,label=$e^+$,tension=0.5}{vr,t4}
\fmf{fermion,label=$e^-$,tension=0.5}{vr,b4}
\fmfdot{vl,v1,v2,vr}
\end{fmfgraph*}}
\nonumber\\[6mm]
&=&{\Green f_{\rho}(m,{\vec p},{\vec x},t)\;
        A_{\rho}(m,{\vec p},{\vec x},t)}\;
2m\;\Gamma^{\rho\;\mbox{\small e}^+\mbox{\small e}^-}(m).
\end{eqnarray}
Here $\Gamma^{\rho\;\mbox{\small e}^+\mbox{\small e}^-}(m)\propto1/m^3$ is the
mass-dependent electromagnetic decay rate of the $\rho$
into the 
di-lepton pair of invariant mass $m$. 
The phase-space distribution $f_{\rho}(m,{\vec p},{\vec
  x},t)$ and the spectral function $A_{\rho}(m,{\vec p},{\vec x},t)$ define
the properties of the $\rho$-meson at space-time point ${\vec x},t$. Both
quantities are in principle to be determined dynamically by an appropriate
transport model. However till to-date the spectral functions are not treated
dynamically in most of the present transport models. Rather one employs
on-shell $\delta$-functions for all stable particles and spectral functions
fixed to the vacuum shape for resonances.

As an illustration, we follow the two channel example presented by one of us
in ref. \cite{Knoll-Erice}. There the $\rho$-meson just strongly
couples to two channels, i.e. the $\pi^+\pi^-$ and $\pi N\leftrightarrow\rho
N$ channels, the latter being relevant at finite nuclear densities. The latter
component is representative for all channels contributing to the so-called
{\em direct $\rho$} in transport codes. For a first orientation the
equilibrium properties\footnote{Far more sophisticated calculations have
already been presented in the
literature~\cite{HFN,Rapp,Mosel,Klingl,FrimanPirner,FLW}.  It is not the point
to compete with them at this place.} are discussed in simple analytical terms
with the aim to discuss the consequences for the implementation of such
resonance processes into dynamical transport simulation codes.

Both considered processes add to the total width of the $\rho$-meson
\begin{eqnarray}\label{Gammatot}
\Gamma_{\rm tot}(m,{\vec p})&=&\Gamma_{\rho\rightarrow{\pi}^+{\pi}^-}(m,{\vec
  p})+ 
\Gamma_{\rho\rightarrow{\pi} NN^{-1}}(m,{\vec p}),
\end{eqnarray}
and the equilibrium spectral function then results from the cuts of the two
diagrams \unitlength1mm
\begin{eqnarray}\label{A2}
{\Green A_{\rho}(m,{\vec p})}\;&=&\normalsize
\underbrace{
\parbox{40mm}{
\begin{fmfgraph*}(40,28)
\fmfpen{thin}
\fmfstraight
\fmftopn{t}{4}
\fmfbottomn{b}{4}
\fmf{phantom,tension=2}{t3,c0}
\fmf{plain,width=0.7}{c0,c1}\fmf{phantom,tension=2}{c1,c2}
\fmf{plain,width=0.7}{c2,c3}
\fmf{phantom,tension=2}{c3,c4}\fmf{plain,width=0.7}{c4,c5}
\fmf{phantom,tension=2}{c5,b2}
\fmfpen{thick}
\fmfleft{l}
\fmfright{r}
\fmf{gluon,label=$\rho$,l.d=15}{lv,l}
\fmf{phantom,tension=0.7}{lv,rv}
\fmf{gluon,label=$\rho$,l.d=15}{r,rv}
\fmffreeze
\fmf{dashes_arrow,left,label=$\pi$}{lv,rv,lv}
\fmfdot{lv,rv}
\end{fmfgraph*}}\;\;+\;\;
\parbox{40mm}{
\begin{fmfgraph*}(40,28)
\fmfpen{thin}
\fmfstraight
\fmftopn{t}{4}
\fmfbottomn{b}{4}
\fmf{phantom,tension=2}{t3,c0}
\fmf{plain,width=0.7}{c0,c1}\fmf{phantom,tension=2}{c1,c2}
\fmf{plain,width=0.7}{c2,c3}
\fmf{phantom,tension=2}{c3,c4}\fmf{plain,width=0.7}{c4,c5}
\fmf{phantom,tension=2}{c5,b2}
\fmfpen{thick}
\fmfleft{l}
\fmfright{r}
\fmf{gluon,label=$\rho$,l.d=15}{lv,l}
\fmf{dashes_arrow,label=$\pi$,tension=0.7}{lv,rv}
\fmf{gluon,label=$\rho$,l.d=15}{r,rv}
\fmffreeze
\fmf{fermion,left,label=$N$}{lv,rv}
\fmf{fermion,left,label=$N^{-1}$}{rv,lv}
\fmfdot{lv,rv}
\end{fmfgraph*}}
}_
{\displaystyle\frac{
        {\Blue2m\Gamma_{\rho\;\pi^+\pi^-}} +
        {\Red2m\Gamma_{\rho\;\pi N N^{-1}}}}
        {\left(m^2-m_\rho^2-\mbox{Re}\Sigma^R\right)^2
        +{\Red m^2\Gamma_{\rm tot}^2}}} .
\end{eqnarray}
In principle, both diagrams have to be calculated in terms of fully
self-consistent propagators, i.e. with corresponding widths for all particles
involved. This formidable task has not been done yet. Using
micro-reversibility and the properties of thermal distributions, the two terms
in Eq. (\ref{A2}) contributing to the di-lepton yield (\ref{dndtdm}), can
indeed approximately be reformulated as the thermal average of a
$\pi^+\pi^-\rightarrow\rho\rightarrow{\rm e}^+{\rm e}^-$-annihilation process
and a $\pi N\rightarrow\rho N\rightarrow{\rm e}^+{\rm e}^-N$-scattering
process, i.e.
\begin{eqnarray}\label{x-sect}
\frac{\di n^{{\rm
  e}^+{\rm e}^-}}{\di m\di t}&\propto&
    \left<f_{\pi^+}f_{\pi^-}\; v_{\pi\pi}\;
      \sigma(\pi^+\pi^-\rightarrow\rho\rightarrow{\rm e}^+{\rm
          e}^-)\vphantom{A^A}\right. \cr&& +\left.
      f_{\pi}f_N\; v_{\pi N}\;\sigma(\pi N\rightarrow\rho N\rightarrow{\rm
          e}^+{\rm e}^-N)\vphantom{A^A}\right>_T ,
\end{eqnarray}
where $f_{\pi}$ and $f_N$ are corresponding particle occupations and
$v_{\pi\pi}$ and $v_{\pi N}$ are relative velocities.
However, the important fact to be noticed is that in order to preserve
unitarity the corresponding cross sections are no longer the free ones, as
given by the vacuum decay width in the denominator, but rather involve the
{\em medium dependent total width} (\ref{Gammatot}). This illustrates in
simple terms that rates of broad resonances can no longer simply be added in a
perturbative way.  Since it concerns a coupled channel problem, there is a
cross talk between the different channels to the extent that the common
resonance propagator attains the total width arising from all partial widths
feeding and depopulating the resonance. While a perturbative treatment with
free cross sections in Eq. (\ref{x-sect}) would enhance the yield at
resonance mass,
$m=m_{\rho}$, if a channel is added, cf. Fig.~3 left part, the correct
treatment (\ref{A2}) even inverts the trend and indeed depletes the yield at
resonance mass, right part in Fig.~3. Furthermore, one sees that only 
the total yield
involves the spectral function, while any partial cross section refers to
that partial term with the corresponding partial width in the numerator!
Compared to the shape of the spectral function both thermal components in
Fig.~3 show a significant enhancement on the low mass side and a strong
depletion at high masses due to the thermal weight $f\propto\exp(-p_0/T)$ in
the rate (\ref{dndtdm}). This kinematical effect related to the broad width
also survives in non-equilibrium calculations and is a signature of phase
space restrictions imposed for particles with higher energies. The related
question how to preserve detailed balance in the case of broad resonances was
addressed by Danielewicz and Bertsch \cite{DB}. The latter method has then
been implemented in transport models mostly applied to the description of the
$\Delta$-resonance. For the transport description of the $\rho$ meson only
quite recently a description level has been realized that properly includes
the width effects discussed above, e.g. in ref. \cite{Effenberger99},
c.f. also the comments given in \cite{Effenberger99-1}. The transport
treatment of broad resonances is further discussed in sections 4 - 7.
\begin{figure}
\unitlength1cm
\begin{picture}(19,6.6)(-2,0)
\put(0,0){\epsfxsize=6cm\epsfbox{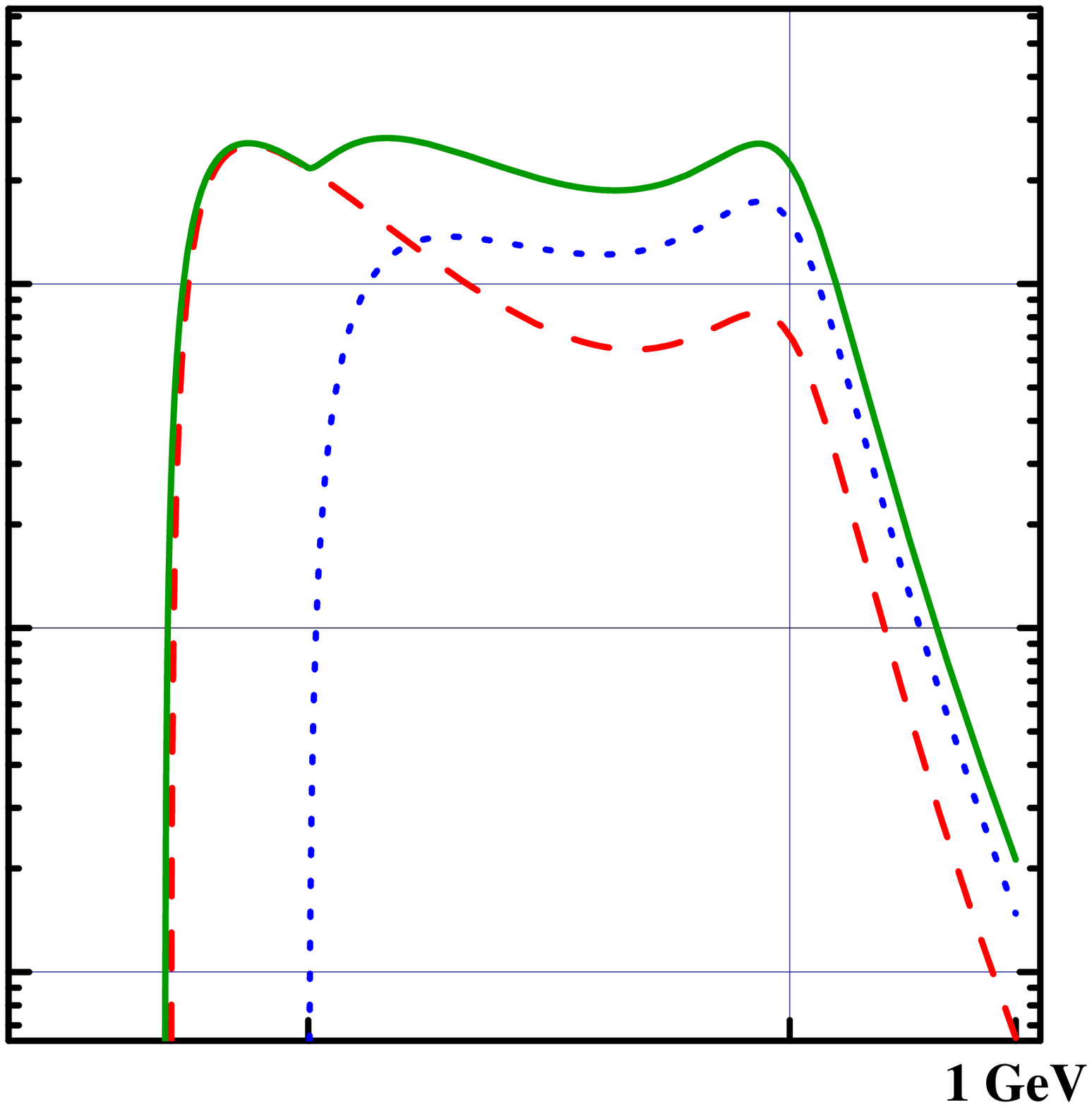}}
\put(6.,0){\epsfxsize=6cm\epsfbox{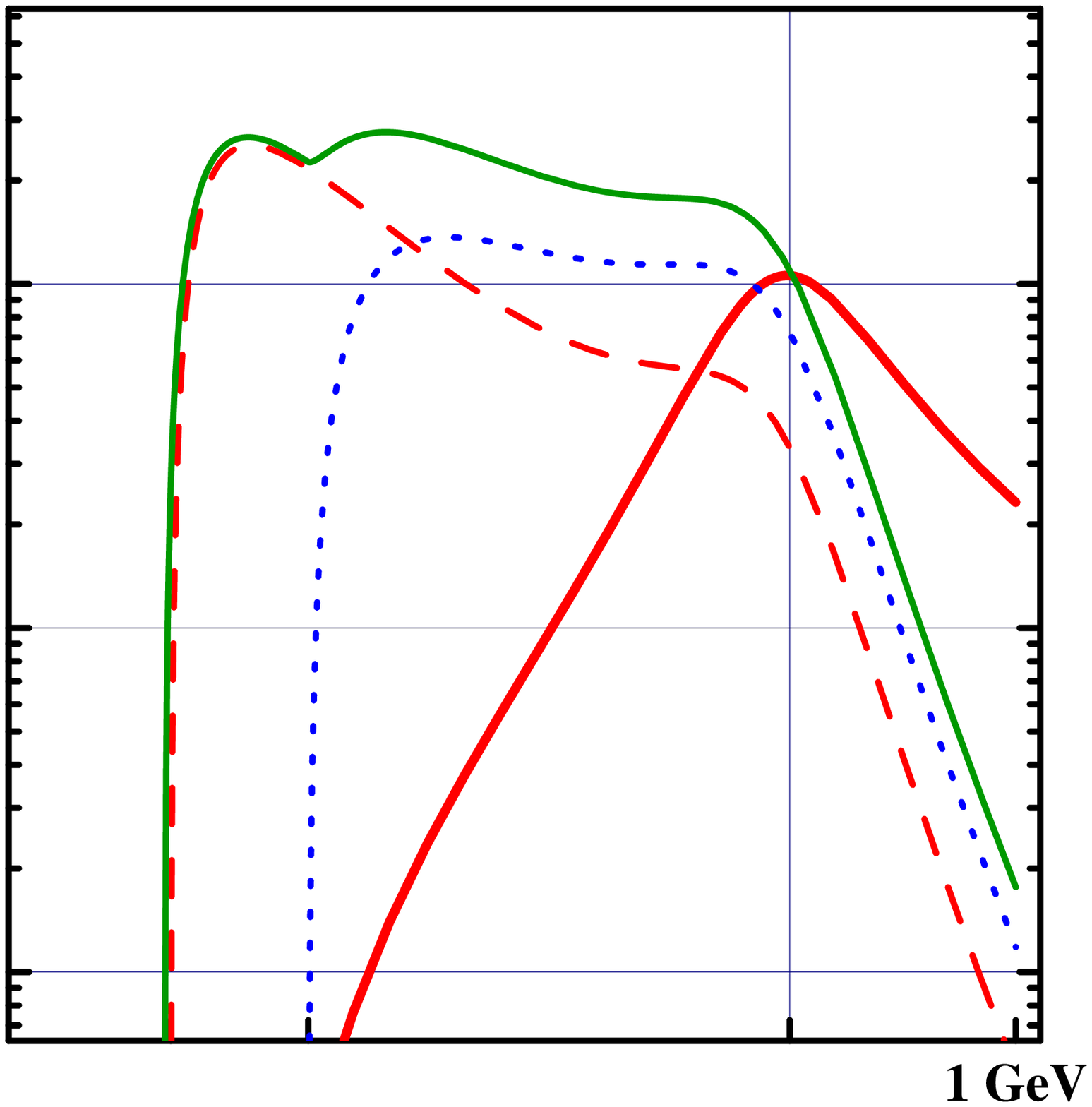}}
\put(6.,6.3){\makebox(0,0){\bf Di-lepton rates from thermal $\rho$-mesons
($T=110$ MeV)}}
\put(1.,0.1){\makebox(0,0){\small$m_{\pi}$}}
\put(1.8,0.1){\makebox(0,0){\small$2m_{\pi}$}}
\put(4.3,0.1){\makebox(0,0){\small$m_{\rho}$}}
\put(6,0){
\put(1.,0.1){\makebox(0,0){\small$m_{\pi}$}}
\put(1.8,0.1){\makebox(0,0){\small$2m_{\pi}$}}
\put(4.3,0.1){\makebox(0,0){\small$m_{\rho}$}}}
\put(3,5.6){\makebox(0,0){$\Gamma_{\rm tot}=\Gamma_{\rm free}$}}
\put(9,5.6){\makebox(0,0){full $\Gamma_{\rm tot}$}}
\put(9.5,3.0){\makebox(0,0){$A_{\rho}$}}
\end{picture}
\caption{
$\mbox{e}^+\mbox{e}^-$ rates (arb.  units)
  as a function of the invariant pair mass $m$ at $T=110$ MeV from
  $\pi^+\pi^-$ annihilation (dotted line) and direct $\rho$-meson contribution
  (dashed line), the full line gives the sum of both contributions. Left part:
  using the free cross section recipe, i.e. with $\Gamma_{\rm
    tot}=\Gamma_{\rho\;\pi^+\pi^-}$; right part: the correct partial rates
  (\ref{A2}). $A_{\rho}$ is given by the thick line. The calculations are done
  with $\Gamma_{\rho\leftrightarrow\pi\pi}(m_{\rho})=150$ MeV and
  $\Gamma_{\rho\leftrightarrow\pi N N^{-1}}(m_{\rho})=70$ MeV.}
\end{figure}
\unitlength=1mm
\begin{figure}
\begin{picture}(118,75)(-25,0)
\put(-37,94){{
\epsfig{file=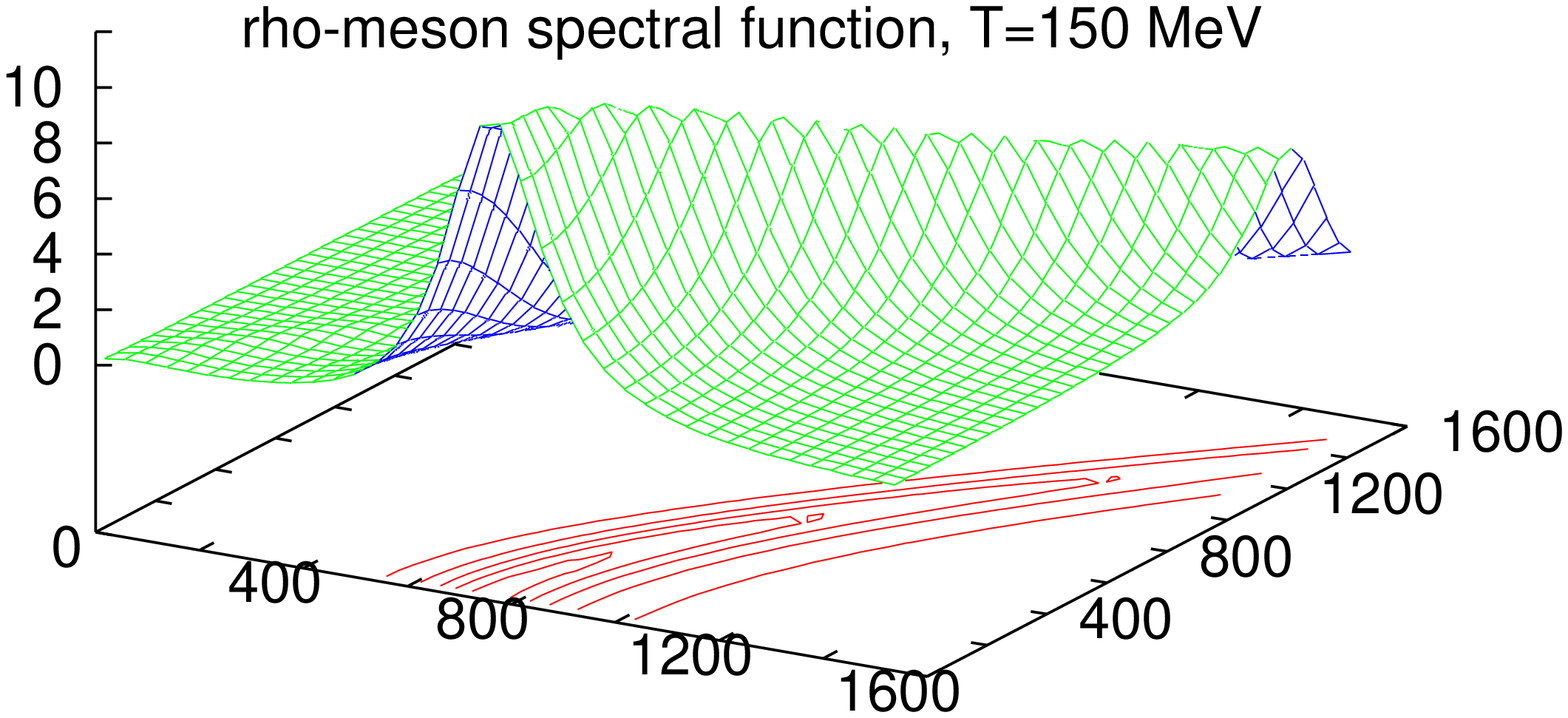,width=10cm,height=10cm,angle=-90}}}
\put(-3,7){$\omega$}
\put(43,12){$|\vec p|$}
\put(55,80){{
\epsfig{file=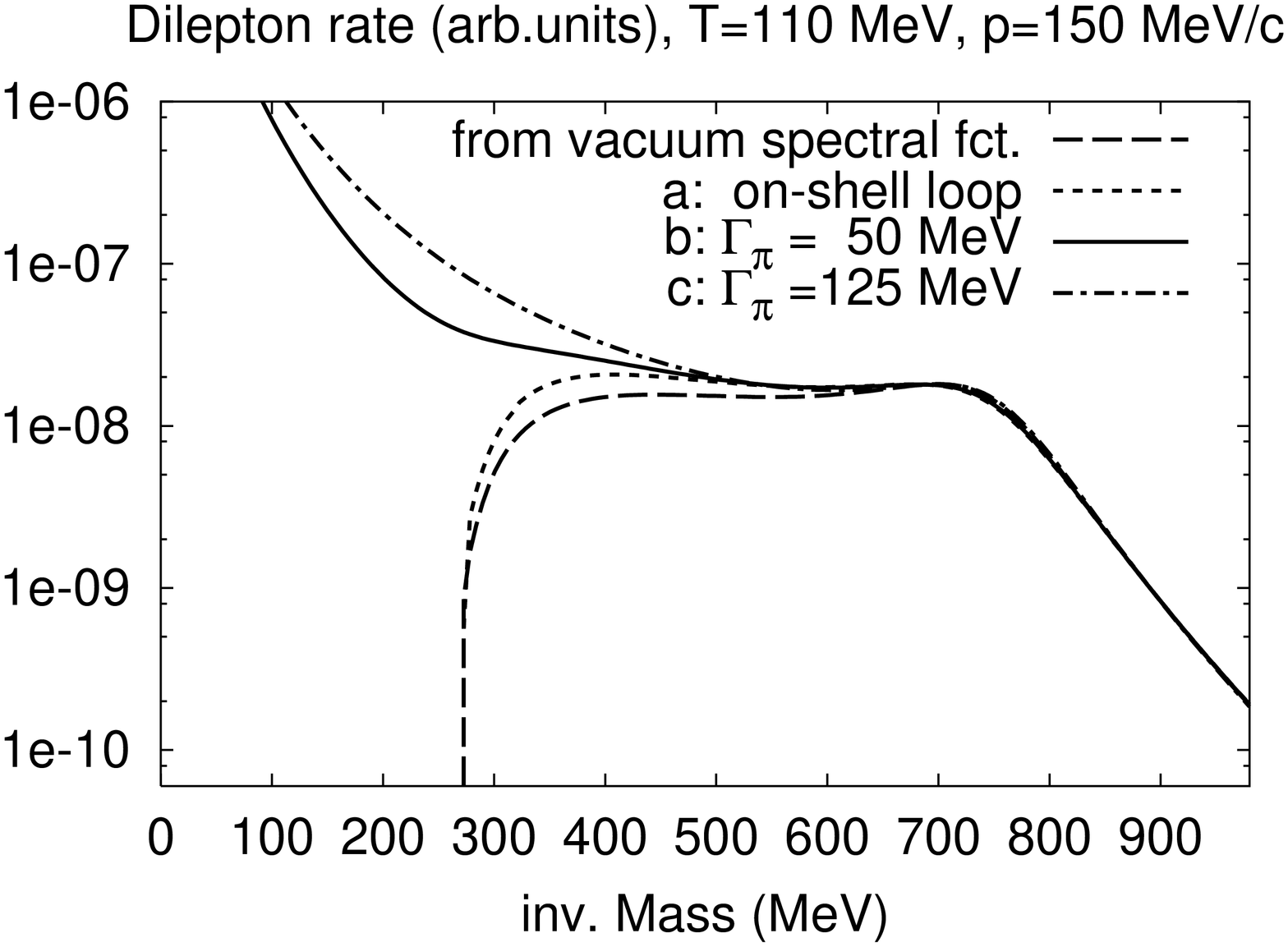,width=8cm,height=8cm,angle=-90}}}
\end{picture}
\caption{Left part: contour plot of the self-consistent
  spectral function of the $\rho$-meson as a function of energy and spatial
  momentum; right part: thermal di-lepton rate as a function of invariant mass
  at $\mid{\vec p}\mid =300$ MeV$/c$}
\end{figure}

As an example we show an exploratory study of the interacting system of $\pi$,
$\rho$ and $a_1$-mesons described by the $\Phi$-functional 
\vspace*{7mm}
\unitlength=1mm
\begin{eqnarray}\label{phi-pi-rho-a1}\cr
\Phi =\;
\parbox{20mm}{
\begin{fmfgraph*}(20,0)
\fmfpen{thick}
\fmfleft{l}
\fmfright{r}
\fmfforce{(0.0w,0.5h)}{l}
\fmfforce{(1.0w,0.5h)}{r}
\fmf{boson,label=$\rho$,label.side=left}{l,r}
\fmf{plain,left=.9,tension=.4,label=\raisebox{-.7mm}{$\pi$}}{l,r}
\fmf{plain,left=.9,tension=.4,label=\raisebox{-.7mm}{$\pi$},l.side=right}{r,l}
\fmfdot{l,r}
\end{fmfgraph*}
}
\;+\;
\parbox{20mm}{
\begin{fmfgraph*}(20,0)
\fmfpen{thick}
\fmfleft{l}
\fmfright{r}
\fmfforce{(0.0w,0.5h)}{l}
\fmfforce{(1.0w,0.5h)}{r}
\fmf{plain,left=.9,tension=.4,label=\raisebox{-.7mm}{$\pi$}}{l,r}
\fmf{boson,label=$\rho$,l.side=left}{l,r}
\fmf{gluon,left=.9,tension=.4,label=\raisebox{1.5mm}{$a_1$},l.side=right}{r,l}
\fmfdot{l,r}
\end{fmfgraph*}
}
\;+\;
\parbox{20mm}{
\begin{fmfgraph*}(20,0)
\fmfpen{thick}
\fmfleft{l}
\fmfright{r}
\fmfforce{(0.0w,0.5h)}{l}
\fmfforce{(1.0w,0.5h)}{r}
\fmf{plain,left=.9,tension=.4,label=\raisebox{-.7mm}{$\pi$},l.side=right}{r,l}
\fmf{plain,left=.3,tension=.4,label=\raisebox{-.7mm}{$\pi$},l.side=right}{r,l}
\fmf{plain,left=.3,tension=.4,label=\raisebox{-.7mm}{$\pi$}}{l,r}
\fmf{plain,left=.9,tension=.4,label=\raisebox{-.7mm}{$\pi$}}{l,r}
\fmfdot{l,r}
\end{fmfgraph*}
}\vphantom{\rule[-12mm]{0mm}{4mm}}
\end{eqnarray}
\noindent (cf.  section
\ref{Phi-derv} below). The couplings and masses are chosen as to reproduce the
known vacuum properties of the $\rho$ and $a_1$ meson with nominal masses and
widths $m_{\rho}=770$ MeV, $m_{a_1}=1200$ MeV, $\Gamma_{\rho}=150$ MeV,
$\Gamma_{a_1}=400$ MeV. The results of a finite temperature calculation at
$T=150$ MeV with all self-energy loops resulting from the $\Phi$-functional of
Eq. (\ref{phi-pi-rho-a1}) computed~\cite{Hees} with self-consistent broad
width Green's 
functions are displayed in Fig. 4 (corrections to the real part of the
self-energies were not yet included). The last diagram of $\Phi$ with the four
pion self-coupling has been added in order to supply pion with broad
mass-width as they would result from the coupling of pions to nucleons and the
$\Delta$ resonance in nuclear matter environment. As compared to first-order
one-loop results which drop to zero below the 2-pion threshold at 280 MeV, the
self-consistent results essentially add in strength at the low-mass side of
the di-lepton spectrum.

\subsection{Virial Limit}\label{Virial limit}

In the dilute-density limit (virial
limit) the corresponding self-energies of the
particles and intermediate resonances are entirely determined by two-body
scattering properties, in particular, by scattering phase shifts. We
illustrate this at the example of the interacting system of nucleons, pions
and delta resonances, which has recently been investigated by Weinhold et al.
\cite{Weinhold}. Following their study we consider a pedagogical example,
where the $\pi NN$-interaction is omitted. Then with a $p$-wave $\pi
N\Delta$-coupling vertex among the three fields the first and only diagram of
$\Phi$ up to two vertices and the corresponding three self-energies are given
by
\begin{eqnarray}\label{piNdelta-diag}\unitlength=1cm
\hspace*{-0.9cm}\mbox{\Large$\Phi$}=
\begin{picture}(2,1)\put(-0.2,1.1){
\includegraphics[width=-2cm,height=2cm,angle=90]{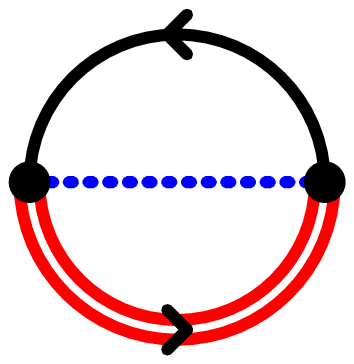}
}
\end{picture}
\hspace*{0.6cm}{\mbox{\large$\Se_{N}$}= }
\begin{picture}(2,1)\put(0.1,1.1){
\includegraphics[width=-2cm,height=2cm,angle=90]{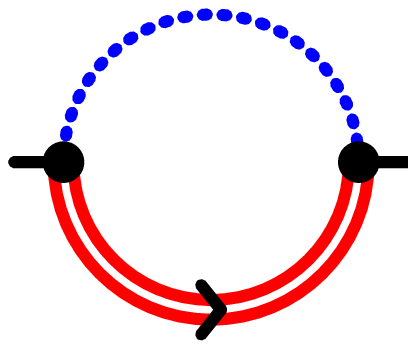}
}
\end{picture}
\hspace*{0.6cm}
{\mbox{\large$\Se_{\pi}$} = }
\begin{picture}(2,1)\put(0.1,1.1){
\includegraphics[width=-2cm,height=2cm,angle=90]{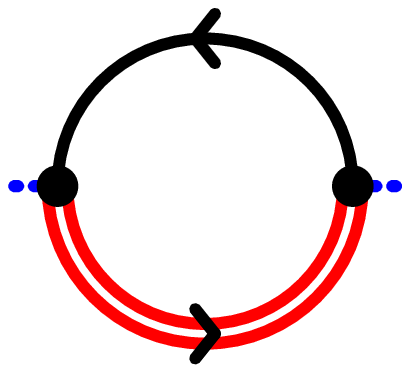}
}
\end{picture}
\hspace*{0.6cm}{\mbox{\large$\Se_{\Delta}$}= }
\begin{picture}(2,1)\put(0.1,1.1){
\includegraphics[width=-2cm,height=2cm,angle=90]{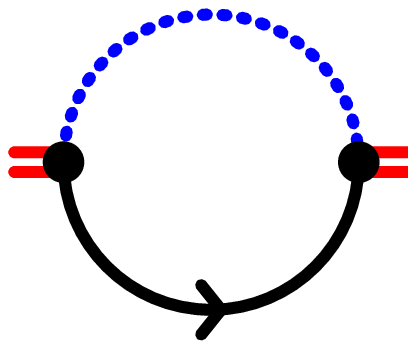}
}
\end{picture}
\\ \nonumber
\end{eqnarray}
Here the solid, dashed and double lines denote the propagators of $N$, $\pi$
and $\Delta$, respectively. In non-relativistic approximation for the baryons
we ignore contributions from the baryon Dirac-sea. Then the bare pion mass
agrees with its vacuum value, while the nucleon and delta masses require
appropriate mass counter terms. The $\Delta$ self-energy $\Se_{\Delta}$
attains the vacuum width and position of the delta resonance due to the decay
into a pion and a nucleon.  The corresponding scattering diagrams are obtained
by opening two propagator lines of $\Phi$ with the prominent feature that the
$\pi N$-scattering proceeds through the delta resonance. Since in this case a
single resonance couples to a single scattering channel, the vacuum spectral
function of the resonance can directly be expressed through the scattering
$T$-matrix and hence through measured scattering phase shifts
\begin{eqnarray}\label{Delta-vac}
\unitlength=1.0cm
\hspace*{-19mm}
\begin{picture}(4,1.5)\put(0.,-.4){
\includegraphics[width=4cm,height=2cm]{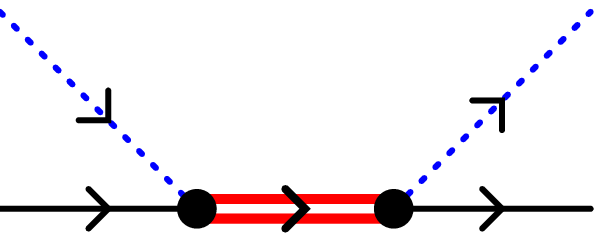}
}
\end{picture}
  &&\quad
  \left|\;T_{33}\;\right|^2 =
  4 \sin^2 \delta_{33}(p)=
  \Gamma_{\Delta}^{\rm vac}(p)A_{\Delta}^{\rm vac}(p),
\end{eqnarray}
where $p=p_N+p_{\pi}$. Thus through (\ref{Delta-vac}) the vacuum properties of
the delta can almost model-independently be obtained from scattering data.

For the multi-component system the renormalized thermodynamic potential
including vacuum counter terms
can be written
as
%
\begin{eqnarray}
\label{Om-maz-piDN} 
&& \hspace*{-1.2cm} 
\Omega\left\{\Gr_\pi, \Gr_N,\Gr_\Delta\right\}
\\\nonumber
&&=
T\;
\sum_{a\in\{\pi,N,\Delta\}} \mp\kappa\Tr \left\{
-\ln\left[-
\Gr^R_a(p_0 +\ii 0 ,\vec p )\right]
+\Gr^R_a \Se^R_a
\right\}_{T,\mu}
+\mbox{\large $\Phi_T$}, 
\end{eqnarray}
\begin{eqnarray}
\label{kappa}
\kappa =
\left\{
\begin{array}{ll}
1/2
\quad&{\mbox{for neutral bosons}},\\[2mm] 
\displaystyle
1 
\quad&{\mbox{for charged bosons and fermions}}.
\end{array}\right.
\end{eqnarray}
For any function $f(p)$ the
thermodynamic trace $\Tr\{...\}_{T,\mu}$ 
is defined as
%
\begin{eqnarray}
\Tr\{f(p)\}_{T,\mu}
=
\mp d \frac{V}{T}
\int \dpi{p} n({p_0-\mu})\;2\;\Im f(p_0+\ii 0,{\vec p})
\end{eqnarray} 
%
an energy integral over thermal occupations
$n(\varepsilon)=\left[\exp(\varepsilon /T)\pm 1\right]^{-1}$, of
Fermi--Dirac/Bose--Einstein type.
The upper sign appears for fermions the lower, for bosons, $d$ is the
degeneracy in that particle channel, and $V$ denotes the volume.  Eq.
(\ref{Om-maz-piDN}) still has the functional property to provide the retarded
Dyson's equations for the $\Gr_a^R$ from the stationary condition which we use
in order to determine the physical value of $\Omega$. For the particular case
here one further can exploit the property
%
\begin{eqnarray*}
\Phi_T=\pm\kappa T\;\Tr\{\Se_a\Gr_a\}_{T,\mu};\quad \mbox{for}\quad
a\in\{N,\pi,\Delta\}\mbox{ and $\Phi$ of form (\ref{piNdelta-diag})},
\end{eqnarray*}
%
valid for $\Phi_T$ that depend linearly on all propagators. Compatible with
the low density limit one can expand the $\Tr \ln \{-\Gr\}$ terms for the
pion and nucleon around the {\em free} propagators, and finally obtains
\begin{eqnarray}
\label{Om-maz-SP} 
\hspace*{-10mm}
\Omega_{\pi N\Delta}
&&=\left.\Omega\left\{\Gr_\pi,\Gr_N,\Gr_\Delta\right\}\vphantom{\sum}
\right|_{\mbox{{\footnotesize stationary}}}\cr
&&=\Omega_N^{\rm free}+\Omega_\pi^{\rm free}
  +\;
T\;   
\Tr \left\{
\ln\left[-\Gr^R_{\Delta}(p_0 +\ii 0 ,\vec p )\right]\right\}_{T,\mu}\\
\label{Omega-phase}
&&=\Omega_N^{\rm free}+\Omega_\pi^{\rm free}
  +d_{\Delta}
T\;
V\int \frac{\di^4 p}{(2\pi)^4}
  2\frac{\partial \delta_{33}(p)}{\partial p_0} 
  \ln \left[1-n_{\Delta}\left(p_0-\mu_{\Delta}\right)\right]
\end{eqnarray} 
%
for the physical {\em value} of $\Omega$. Here the $\Omega_a^{\rm free}$ are
the {\em free} single-particle thermodynamic potentials\footnote{The
appropriate cancellation of terms for the result (\ref{Om-maz-SP}) is only
achieved, if one uses $\Omega^{\rm free}$, i.e. the partition sum of free
particles with the free energy--momentum dispersion relation. Within this
model already on the {\em vacuum} level the nucleon would acquire loop
corrections to its self-energy which would lead to deviations between
$\Omega^{\rm vac}$ and $\Omega^{\rm free}$, as well as between the
corresponding propagators off their mass shell.}, while
$\mu_{\Delta}=\mu_N+\mu_{\pi}$ and $d_{\Delta}=16$ are the chemical potential
and degeneracy factor of the $\Delta$ resonance, respectively.  The last term
in (\ref{Omega-phase}) obtained through (\ref{Delta-vac}), represents a famous
result derived by Beth-Uhlenbeck \cite{BethU,Huang}, later generalized by
Dashen, Ma and Bernstein \cite{DMB} and applied to nuclear resonance matter in
refs.  \cite{Mekjian,VPrakash,Weinhold}.  It illustrates that the virial
corrections of the system's level density due to interactions are entirely
given by the energy variation of the corresponding two-body scattering phase
shifts $\partial \delta/\partial p_0$.

All thermodynamic properties can be obtained from $\Omega$ through partial
differentiations with respect to $T$ and the $\mu$. The final form
(\ref{Omega-phase}) may give the impression that one deals with
non-interacting nucleons and pions. This is however not the case.  For
instance the densities of baryons and pions derived from (\ref{Omega-phase}) 
become\footnote{In equilibrium $\mu_{\pi}$ has to
  be put to zero after differentiation.} 
%
\begin{eqnarray}
\hspace*{-6mm}
\rho_B =\frac{\partial\Omega_{\pi N\Delta}}{\partial\mu_N}
   =\rho_N^{\rm free}+\rho_{\Delta}+\rho_{\rm corr},
\,\,\,\,
\rho_{\pi}=\frac{\partial\Omega_{\pi N\Delta}}{\partial\mu_{\pi}}
   =\rho_{\pi}^{\rm free}+\rho_{\Delta}+\rho_{\rm corr},
\end{eqnarray}
%
with
\vspace*{-6mm}
%
\begin{eqnarray}
\rho_{\Delta}&=&
   d_{\Delta}\int\frac{\di^4p}{(2\pi)^4}
   n_{\Delta}(p_0-\mu_{\Delta})A^{\rm vac}_{\Delta}(p),
\\
\rho_{\rm corr}&=&
   d_{\Delta}\int\frac{\di^4p}{(2\pi)^4}
   n_{\Delta}(p_0-\mu_{\Delta})B_{\rm corr}(p),
\end{eqnarray}
%
Here the density of deltas
$\rho_{\Delta}$ is determined by the delta spectral function. The interaction
contribution contained in the correlation density $\rho_{\rm corr}$ depends on
the difference between the phase-shift variation and the spectral function
%
\begin{eqnarray}
B_{\rm corr}&=&2\frac{\partial\delta_{33}(p_0 )}{\partial p_0 }
   -A^{\rm vac}_{\Delta}(p)
   =2\Im\left[\frac{\partial\Se_{\Delta}^R (p)}
     {\partial p_0 }{\Gr^{\rm vac}_{\Delta}}^R (p)\right] .
\end{eqnarray}
%
Due to the fact that $\Gamma_{\Delta}(p)$ grows with energy and the real part
of 
$\Gr_{\Delta}$ changes sign at the resonance energy, $B_{\rm corr}$ becomes
positive below and negative above resonance, respectively. It leads to an
enhancement of both densities at low energies, i.e. below resonance and this
way to a further softening of the resulting equation of state compared to the
naive spectral function treatment ignoring the $B_{\rm corr}$ terms.  This
illustrates that an interacting resonance gas {\em cannot} consistently be
described by a set of free particles (here the pions and nucleons) plus vacuum
resonances (here the delta), described by their spectral function. Rather the
coupling of a bare resonance to the stable particles determines its width, and
thus its spectral properties in vacuum.  At the same time the stable particles
are modified due to the interaction with the resonance.  Only the account of
all three self-energies in (\ref{piNdelta-diag}) provides a conserving and
thermodynamically consistent approximation.

Alternatively to the picture above, the properties of the system can be
discussed entirely in terms of the asymptotic particles, i.e. the pion and the
nucleon. The thermodynamic potential is then still given by
(\ref{Omega-phase}). This form is valid even without intermediate resonances
and the phase-shifts just account for the $\pi N$ interaction properties. Also
the self-energy of the pion can be obtained from phase shifts by means of the
optical theorem \cite{Lenz,Land77}. To linear order in the nucleon density
$\rho_N$ one determines the pion self-energy

\begin{eqnarray}\label{pi-opt}
\Se_{\pi}(p_{\rm lab})=4\pi\frac{p_{\rm lab}}{p_{\rm cm}} \rho_N F_{\pi N}(0)
   =-2\pi \frac{ p_{\rm lab}}{p_{\rm
   cm}^2}\frac{d_{\Delta}}{d_{\pi}}\;\frac{\rho_N}{d_N}  
   2\sin\delta_{33}e^{\ii \delta_{33}},
\end{eqnarray}
%
from the forward $\pi N$-scattering amplitude $F_{\pi N}(0)$. Here $p_{\rm
lab}$ and $p_{\rm cm}$ refer to the pion 3-momenta in the matter rest frame
and the c.m. frame of the $\pi N$ coliisions. The arising kinematical factor
${p_{\rm lab}}/{p_{\rm cm}}=\sqrt{s}/m_N$ 
which has mostly escaped notice even in standard
references on the low desity theorem, e.g. \cite{Dover71}, becomes important
for heavier projectiles like kaons, cf.  ref. \cite{Lutz94}. 
Here the degeneracy factors
$d_N:d_{\pi}:d_{\Delta}=4:3:16$ just provide the proper spin/isospin
counting. This self-energy, which determines an optical potential or index of
refraction, is attractive below the delta resonance energy and repulsive
above. It agrees with a related effect in optics, where a resonance in the
medium causes an anomalous behavior of the real part of the index of
refraction, which is larger than 1 below the resonance frequency and less than
1 above the resonance.  Thus, absorption, e.g. by exciting a resonance, is
always accompanied by a change of the real part of the index of refraction of
the scattered particle. The $\Phi$-derivable principle automatically takes
care about these features.

As has been discussed in \cite{DP}, the corrections to the system's level
density (last term in (\ref{Omega-phase})) can also be inferred from the time
shifts (or time delays) induced by the scattering processes.  From ergodicity
arguments \cite{DP} one obtains for a single partial wave
%
\begin{eqnarray}
\frac{\partial}{\partial p_0} 
&&\left(N_{\rm level}(p_0 )-N_{\rm level}^{\rm free}(p_0 )\right)
   =\tau_{\rm forward}+\tau_{\rm scatt.}=\tau_{\rm delay}\cr
  &&=2\frac{\partial}{\partial p_0 }\left[\sin\delta_{33}\cos\delta_{33}\right]
   +4\sin^2\delta_{33}\frac{\partial \delta_{33}}{\partial p_0}
   =2\frac{\partial \delta_{33}}{\partial p_0}.
\end{eqnarray}
%
These expressions apply to the c.m. frame.  Here the forward delay time
$\tau_{\rm forward}$ relates to the change of the group velocity
induced by the real part of the optical potential, c.f.  (\ref{pi-opt}). The
scattering time $\tau_{\rm scatt.}$ finally results from the delayed
re-emission of the pion from the intermediate resonance to angles off the
forward direction.
\section{Quantum Kinetic Equation}

The three above-presented examples unambiguously show that for a
consistent dynamical treatment of non-equilibrium evolution of soft
radiation and broad resonances we need a transport theory that takes
due account of mass-widths of constituent particles. A proper frame
for such a transport is provided by Kadanoff--Baym equations. We
consider the Kadanoff--Baym equations 
in the first-order gradient approximation,
assuming that time--space evolution of a system is smooth enough to
justify this approximation. 

First of all, it is helpful to avoid all the imaginary factors inherent in the
standard Green's function formulation ($\Gr^{ij}$ with $i,j\in\{-+\}$) and
introduce quantities which are real and, in the quasi-homogeneous limit,
positive and therefore have a straightforward physical interpretation 
\cite{IKV99}, much
like for the Boltzmann equation.  In the Wigner representation we define
%
\begin{eqnarray}
\label{F}
\Fd (X,p) &=& \A (X,p) \fd (X,p)
 =  (\mp )\ii \Gr^{-+} (X,p) , \nonumber\\
\Fdt (X,p) &=& \A (X,p) [1 \mp \fd (X,p)] = \ii \Gr^{+-} (X,p), \\ 
\label{A}
 A (X,p) &\equiv& -2\Im \Gr^R (X,p) = \Fdt \pm \Fd =
\ii \left(\Ga^{+-}-\Ga^{-+}\right)
\end{eqnarray}
%
for the generalized Wigner functions $\F$ and $\Ft$ with the corresponding
{\em four}-phase-space distribution functions $\fd(X,p)$ and Fermi/Bose
factors $[1 \mp \fd (X,p)]$, with the spectral function $A(X,p)$ and the
retarded 
propagator $\Gr^R$.  Here and below the upper sign corresponds to fermions and
the lower one, to bosons. According to relations between Green's functions
$\Gr^{ij}$ {\em only two independent real functions of all the $\Gr^{ij}$ are
  required for a complete description}.  Likewise the reduced gain and loss
rates of the collision integral and the damping rate are defined as
%
\def\ga{\gamma}
\begin{eqnarray}
\label{gain}
\Ldt (X,p) 
&=&   (\mp )\ii \Se^{-+} (X,p), \quad
\Ld (X,p)  
=  \ii \Se^{+-} (X,p), \\
\label{G-def}
\Gamma (X,p)&\equiv& -2\Im \Se^R (X,p) = \Ld (X,p)\pm\Ldt (X,p), 
\end{eqnarray}
%
where $\Se^{ij}$  are contour components of the self-energy, and
$\Se^R$ is the retarded self-energy. 
 
In terms of this notation and within the
first-order gradient approximation, the Kadanoff--Baym equations 
for $\Fd$ and $\Fdt$ (which result from differences of the 
corresponding Dyson's equations with their adjoint ones) take the 
kinetic form 
%
\begin{eqnarray} 
\label{keqk1} 
{\cal D}\Fd  -  
\Pbr{\Ldt , 
\Re\Gr^R} &=& C  , 
\\
\label{keqkt1} 
{\cal D}\Fdt - 
\Pbr{\Ld , \Re\Gr^R} &=&\mp C  
\end{eqnarray} 
%
with 
the drift operator and collision term respectively
%
\begin{eqnarray}
\label{Coll(kin)}
{\cal D}&=&\left(2\pi_{\mu}
- 
\frac{\partial \Re\Sa^R}{\partial p^{\mu}} 
\right) 
\partial^{\mu}_X + 
\frac{\partial \Re\Sa^R}{\partial X^{\mu}}  
\frac{\partial }{\partial p_{\mu}}, 
\\
C (X,p) &=&
\Ldt (X,p) \Ft (X,p) 
- \Ld (X,p) \F (X,p).
\end{eqnarray}
%
$2\pi_{\mu}= v^\mu=(1,\vec{p}/m)$ for non-relativistic particles and 
$\pi_{\mu}=p_{\mu}$ for relativistic bosons.
Within the same approximation level there are two
alternative equations for $\Fd$ and $\Fdt$
%
\begin{eqnarray}
\label{mseq(k)1}
M\Fd - \Re\Ga^R\Ldt
&=&\frac{1}{4}\left(\Pbr{\Gm,\Fd} - \Pbr{\Ldt,\A}\right),
\\\label{mseqt(k)1}
M\Fdt - \Re\Ga^R\Ld
&=&\frac{1}{4}\left(\Pbr{\Gm,\Fdt} - \Pbr{\Ld,\A}\right)
\end{eqnarray}
%
with the  ``mass'' function 
%
\begin{eqnarray}
\label{m-funct}
M(X,p)=
\left\{
\begin{array}{ll}
\displaystyle
p_0 -\vec{p}^2/2m -\Re\Se^R (X,p)  
\quad&{\mbox{for non-relativistic particles}},\\[2mm] 
\displaystyle
-m^2 + p^2 -\Re\Se^R (X,p)
\quad&{\mbox{for relativistic bosons}} 
\end{array}\right.
\end{eqnarray}
%
These two equations result from sums of the corresponding Dyson's equations 
with their adjoint ones. Eqs. (\ref{mseq(k)1}) and (\ref{mseqt(k)1})
can be called the mass-shell equations, 
since in the quasiparticle limit they provide the on-mass-shell condition
$M=0$. 
Appropriate combinations of the two sets
(\ref{keqk1})--(\ref{keqkt1}) and (\ref{mseq(k)1})--(\ref{mseqt(k)1}) provide
us with retarded Green's function equations, 
which are simultaneously solved~\cite{Kad62,BM} by
%
\begin{eqnarray}
\label{Asol}
\Gr^R=\frac{1}{M(X,p)+\ii\Gamma(X,p)/2}\Rightarrow
\left\{\begin{array}{rcl}
A &=&\displaystyle
\frac{\Gamma}{M^2 + \Gamma^2 /4},\\[2mm]
\Re\Gr^R &=& \displaystyle 
\frac{M}{M^2 + \Gamma^2 /4}.
\end{array}\right. 
\end{eqnarray}
%

With the solution (\ref{Asol}) for $\Gr^R$ equations (\ref{keqk1}) and
(\ref{keqkt1}) become identical to each other, as well as Eqs.
(\ref{mseq(k)1}) and (\ref{mseqt(k)1}). However, Eqs.
(\ref{keqk1})--(\ref{keqkt1}) still are not identical to Eqs.
(\ref{mseq(k)1})--(\ref{mseqt(k)1}), while they were identical before the
gradient expansion. Indeed, one can show~\cite{IKV99} that Eqs.
(\ref{keqk1})--(\ref{keqkt1}) differ from Eqs.
(\ref{mseq(k)1})--(\ref{mseqt(k)1}) in second order gradient terms. This is 
acceptable within the gradient approximation, however, the still remaining
difference in the second-order terms is inconvenient from the practical point
of view.  Following Botermans and Malfliet~\cite{BM}, we express $\Ldt=\Gm
f+O(\partial_X)$ and $\Ldt=\Gm (1\mp f)+O(\partial_X)$ from the l.h.s. of
mass-shell Eqs.  (\ref{mseq(k)1}) and (\ref{mseqt(k)1}), substitute them into
the Poisson bracketed terms of Eqs. (\ref{keqk1}) and (\ref{keqkt1}), and
neglect all the resulting second-order gradient terms.  The so obtained {\em
  quantum four-phase-space kinetic equations for $\Fd=fA$ and $\Fdt=(1\mp f)
  A$} then read
%
\begin{eqnarray}
\label{keqk}
\Do 
\left(f A\right) - 
\Pbr{\Gm f,\Re\Gr^R} &=& C , 
\\
\label{keqkt}
\Do 
\left((1\mp f) A\right) - 
\Pbr{\Gm (1\mp f),\Re\Gr^R} &=& \mp C .  
\end{eqnarray}
%
These quantum four-phase-space kinetic equations, which are identical to each
other in view of the retarded relation (\ref{Asol}), are at the same time
completely identical to the correspondingly substituted mass-shell Eqs.
(\ref{mseq(k)1}) and (\ref{mseqt(k)1}). 

The validity of the gradient approximation~\cite{IKV99} relies on the overall
smallness of the collision term $C=\{\mbox{gain} - \mbox{loss}\}$ rather than
on the smallness of the damping width $\Gamma$. Indeed, while fluctuations and
correlations are governed by time scales given by $\Gamma$, the Kadanoff--Baym
equations describe the behavior of the ensemble mean of the occupation in
phase-space $F(X,p)$. It implies that $F(X,p)$ varies on space-time scales
determined by $C$. In cases where $\Gamma$ is not small enough by itself, the
system has to be sufficiently close to equilibrium in order to provide a valid
gradient approximation through the smallness of the collision term $C$.
Both the Kadanoff--Baym (KB), eq. (\ref{keqk1}), and the Botermans--Malfliet
(BM) choice (\ref{keqk}) are, of course, equivalent within the validity range
of the first-order gradient approximation. Frequently, however, such equations
are used beyond the limits of their validity as ad-hoc equations, and then the
different versions may lead to different results. So far we have no physical
condition to prefer one of the choices. The procedure, where in all Poisson
brackets the $\Ldt$ and $\Ld$ terms have consistently been replaced by
$f\Gamma$ and $(1\mp f)\Gamma$, respectively, is therefore optional. However,
in doing so we gained some advantages. Beside the fact that quantum
four-phase-space kinetic equation (\ref{keqk}) and the mass-shell equation are
then {\em exactly} equivalent to each other, this set of equations has a
particular features with respect to the definition of a non-equilibrium
entropy flow in connection with the formulation of an {\em exact} H-theorem in
certain cases.  If we omit these substitutions, both these features would
become approximate with deviations at the second-order gradient level.
A numerical scheme  of the BM choice in application to heavy ion collisions
is constructed in
refs. \cite{Leupold00,Juchem00}.


The equations so far presented, mostly with the Kadanoff--Baym choice
(\ref{keqk1}), were the starting point for many derivations of
extended Boltzmann and generalized kinetic equations, ever since these
equations have been formulated in 1962.  Most of those derivations use
the equal-time reduction by integrating the four-phase-space equations
over energy $p_0$, thus reducing the description to three-phase-space
information, cf. 
refs.~\cite{Bez,LSV,LipS,Kraft,Bonitz,Bornath,SCFNW,Jeon,VBRS} and
refs.  therein. This can only consistently be done in the limit of small width
$\Gamma$ employing some kind of quasi-particle ansatz for the spectral
function $A(X,p)$. Particular attention has been payed to the treatment of the
time-derivative parts in the Poisson brackets, which in the four-phase-space
formulation still appear time-local, i.e. Markovian, while they lead to
retardation effects in the equal-time reduction. Generalized quasiparticle
ans\"atze were proposed, which essentially improve the quality and consistency
of the approximation, providing those extra terms to the naive Boltzmann
equation (some times called additional collision term) which are responsible
for the correct second-order virial corrections and the appropriate
conservation of total energy, cf. \cite{LipS,Bornath} and refs. therein.
However, all these derivations imply some information loss about the
differential mass spectrum due to the inherent reduction to a 3-momentum
representation of the distribution functions by some specific ansatz.  With
the aim to treat cases as those displayed in Figs. 2 and 3, where the
differential mass spectrum can be observed by di-lepton spectra, within a
self-consistent non-equilibrium approach, one has to treat the differential
mass information dynamically, i.e. by means of Eq. (\ref{Asol}) avoiding any
kind of quasi-particle reductions and work with the full quantum
four phase-space kinetic Eq.  (\ref{keqk}). In the following we discuss the
properties of this set of quantum kinetic equations.


\section{$\Phi$-Derivable Approximations 
}\label{Phi-derv}

The preceding considerations have shown that one needs a transport scheme
adapted to broad resonances. Besides the conservation laws it should comply
with requirements of unitarity and detailed balance. A practical suggestion
has been given in ref.~\cite{DB} in terms of cross-sections. However, this
picture is tied to the concept of asymptotic states and therefore not well
suited for the general case, in particular, if more than one channel feeds
into a broad resonance. Therefore, we suggest to revive the so-called
$\Phi$-derivable scheme, originally proposed by Baym~\cite{Baym} on the basis
of the generating functional, or partition sum, given by Luttinger and
Ward~\cite{Luttinger}, and later reformulated in terms of
path-integrals~\cite{Cornwall}.

With the aim to come to a self-consistent and conserving treatment on the
two-point function level, we generalized the $\Phi$-functional method
\cite{Luttinger,Baym} to the real-time contour (${\cal C}$) in ref. 
\cite{IKV}. It was based on a decomposition of the generating functional
$\Gamma$ with bilocal sources into a two-particle reducible part and an
auxiliary functional $\Phi$ which compiles all two-particle-irreducible (2PI)
vacuum diagrams \unitlength=.8cm
%
\begin{eqnarray}\label{keediag}
\hspace*{-5mm}
\ii\Gamma\left\{\Gr , \phi, \lambda \right\} &=& \ii
\Gamma^0\left\{\Gr^0\right\} + \oint \di x \Lg^0\{\phi,\partial_\mu\phi\}
\nonumber 
\\
&+&
\left\{\vhight{1.6}\right.
\underbrace{\sum_{n_\Se}\vhight{1.6}\frac{1}{n_\Se}\GlnG0Sa}
_{\displaystyle \pm \ln\left(1-\odot\Gr^{0}\odot\Se\right)}
\underbrace{-\vhight{1.6}\GGaSa}
_{\displaystyle \pm \odot\Gr\odot\Se\vphantom{\left(\Ga^{0}\right)}}
\left.\vhight{1.6}\right\}
\underbrace{+\vhight{1.6}\sum_{n_\lambda}\frac{1}{n_\lambda}
\Dclosed{\mbox{\scriptsize 2PI}}{\thicklines}}
_{\displaystyle\vphantom{\left(\Ga^{0}\right)} +
\ii\Phi\left\{\Gr , \phi,\lambda \right\}}.
\end{eqnarray}
%
Here $ {\cal L}^0(\phi)$ is the free classical Lagrangian of the
classical field $\phi$, $\Gr^{0}$ and $\Gr$ denote the free and full
contour Green's functions, while $\Se$ is the full contour self-energy of
the particles.  Contrary to the perturbation theory, here the auxiliary
functional $\Phi$ is given by all two particle irreducible closed
diagrams in terms of {\em full} propagators $\Gr$, {\em full} time
dependent classical fields $\phi$ and bare vertices. 
Upper signs in Eq. (\ref{keediag}) relate to fermion quantities, whereas lower
signs, to boson 
ones, while $n_\Se$ and $n_\lambda$ count the number of self-energy
insertions in the ring diagrams and the number of vertices in the diagrams of
$\Phi$, respectively, $\lambda$ is the scaling factor in each vertex. The
stationarity conditions 
%
\begin{eqnarray}
\label{varG/phi}
\delta \Gamma \left\{\Gr , \phi,\lambda \right\}/ \delta \Gr = 0, \quad\quad 
\delta \Gamma \left\{\Gr , \phi,\lambda \right\}/ \delta \phi = 0
\end{eqnarray}
%
provide the set of coupled
equations of motion for the classical fields $\phi$ and Green's
functions $\Gr$ (Dyson Eq.)
\begin{eqnarray}\label{fields}
\phi(x)&=&\phi^0(x) -\oint\di y\Ga^{0}(x,y) J(y),\\ 
\label{Dyson}
\Gr(x,y)&=&\Gr^0(x,y)+\oint\di z\di z'\Gr^0(x,z)\Se(z,z')\Gr(z',y), 
\end{eqnarray}
where the superscript $^0$ marks the free Green's functions and classical
fields. 
The functional $\Phi\{\Gr,\phi\}$ acts as the generating functional for the
self-energy $\Se$ and source currents $J(x)$ via the functional variations
%
\begin{eqnarray}\label{varphi}
\ii J(x)=\frac{\delta\ii \Phi}{\delta \phi(x)},
\quad\quad
-\ii\Se(x,y)=\frac{\delta\ii \Phi}{\delta \ii\Gr(y,x)}.
\end{eqnarray}
%

The advantage of this formulation is that $\Phi$ can be truncated at
any level, thus defining approximation schemes with built in internal
consistency with respect to conservation laws and thermodynamic
consistency. For details we refer to 
\cite{Baym,Luttinger} and our previous paper \cite{IKV}. Note that $\Phi$
itself is constructed in terms of 
``full'' Green's functions, where ``full'' now takes the sense of
solving self-consistently the Dyson's equation with the driving term
derived from this approximate $\Phi$ through relation
(\ref{varphi}). It means that even restricting ourselves to a single
diagram in $\Phi$, in fact, we deal with a whole sub-series of
diagrams in terms of free propagators, and ``full'' takes the sense of
the sum of this whole sub-series. Thus
restricting the infinite set of diagrams for $\Phi$ to either only a
few of them or some sub-series of them defines a $\Phi$-derivable
approximation. Such approximations 
have the following distinct properties: (a) they are conserving, if
$\Phi$ preserves the invariances and symmetries of the Lagrangian for
the full theory; (b) lead to a consistent dynamics, and (c) are
thermodynamically consistent.  
These properties originally shown within the imaginary time formalism with a
time-dependent external perturbation \cite{Baym,Luttinger} also to 
hold in the genuine
non-equilibrium case formulated in the real-time field theory \cite{IKV}.

Transport equation (\ref{keqk}) weighted either with the charge $e$ or with
4-momentum $p^\nu$, integrated over momentum and summed over internal degrees
of freedom like spin ($\Tr$) gives rise to the charge or energy--momentum
conservation laws, respectively, with the Noether 4-current and Noether
energy--momentum tensor defined by the following expressions \cite{IKV99}
%
\begin{eqnarray}
\label{c-new-currentk} 
j^{\mu} (X) 
&=& e \mbox{Tr} \int \dpi{p}
2\pi^{\mu} 
\Fd (X,p), \\
\label{E-M-new-tensork}
\Theta^{\mu\nu}(X)
&=&
\mbox{Tr} \int \dpi{p} 
2\pi^{\mu} p^{\nu} \Fd (X,p)
+ g^{\mu\nu}\left(
{\cal E}^{\scr{int}}(X)-{\cal E}^{\scr{pot}}(X)
\right).  
\end{eqnarray}
%
Here 
%
\begin{eqnarray}
\label{eps-int} 
{\cal E}^{\scr{int}}(X)=\left<-\Lint(X)\right>
=\left.\frac{\delta\Phi}{\delta\lambda(x)}\right|_{\lambda=1}
\end{eqnarray}
%
is interaction energy density, which in terms of $\Phi$ is given by a
functional variation with respect to a space-time dependent coupling strength
of interaction part of the Lagrangian density 
$\Lint\rightarrow\lambda(x)\Lint$, cf. ref. \cite{IKV}.  The potential
energy density ${\cal E}^{\scr{pot}}$ takes the form
%
\begin{eqnarray}
\label{eps-potk}
{\cal E}^{\scr{pot}}
= 
\mbox{Tr}
\int\dpi{p} \left[
\Re\Sa^R \Fd
+ \Re\Ga^R \frac{\Gm}{\A}\F
\right].  
\end{eqnarray}
%
Whereas the first term in squared brackets complies with quasiparticle
expectations, namely mean potential times density, the second term displays
the role of fluctuations in the potential energy density.

The conservation laws only hold, if all the self-energies are
$\Phi$-derivable. In ref. \cite{IKV99}, it was shown that this implies the
following consistency relations (a) for the conserved current
%
\begin{equation}
\label{invarJk}
\ii \mbox{Tr} \int \dpi{p} 
\left[
\Pbr{\Re\Sa^R,
\Fd} 
- 
\Pbr{\Re\Ga^R,\frac{\Gm}{\A}
\F}  
+ C
\right]
=0, 
\end{equation}
%
and (b) for the energy-momentum tensor
%
\begin{eqnarray}
\label{epsilon-invk}
\partial^{\nu}
\left(
{\cal E}^{\scr{pot}} - {\cal E}^{\scr{int}}
\right)
= \mbox{Tr}
\int 
\frac{p^\nu \di^4 p}{(2\pi )^4}
\left[
\Pbr{\Re\Sa^R,
\Fd} 
- 
\Pbr{\Re\Ga^R,\frac{\Gm}{\A}
\F}  
+C
\right] .
\end{eqnarray}
%
They are obtained after
first-order gradient expansion of the corresponding exact relations. 
The contributions from the Markovian collision term $C$ drop out in both
cases, cf. Eq. (\ref{Multi-rate}) below.  The first term in each of the two
relations refers to the change from the free velocity $\vec v$ to the group
velocity $\vec v_g$ in the medium. It can therefore be
associated with a corresponding {\em drag--flow} contribution of the
surrounding matter to the current or energy--momentum flow. The second
(fluctuation) term compensates the former contribution and can therefore be
associated with a {\em back--flow} contribution, which restores the Noether
expressions (\ref{c-new-currentk}) and (\ref{E-M-new-tensork}) to be indeed
the conserved quantities. In this compensation we see the essential role of
fluctuations in the quantum kinetic description.
Dropping or approximating this term would spoil the conservation laws. Indeed,
both expressions (\ref{c-new-currentk}) and (\ref{E-M-new-tensork}) comply
with the quantum kinetic equation (\ref{keqk}), being approximate (up to the
first-order gradient terms) integrals of it.
 
Expressions (\ref{c-new-currentk}) and (\ref{E-M-new-tensork}) for 4-current
end energy--momentum tensor, respectively, as well as self-consistency
relations (\ref{invarJk}) and (\ref{epsilon-invk}) still need a 
renormalization.
They are written explicitly for
the case of non-relativistic particles which number is conserved. This follows
from the conventional way of non-relativistic renormalization for such
particles based on normal ordering. When the number of particles is not
conserved (e.g., for phonons) or a system of relativistic particles is
considered, one should replace $\Fd (X,p) \rightarrow \frac{1}{2}\left(\Fd
  (X,p) \mp \Fdt (X,p) \right)$ in all above formulas in order to take proper
account of zero point vibrations (e.g., of phonons) or of the vacuum
polarization in the relativistic case.  These symmetrized equations, derived
from special ($\mp$) combinations of the transport equations (\ref{keqk}) and
(\ref{keqkt}), are generally ultra-violet divergent, and hence, have to be
properly renormalized at the vacuum level.

\section{Collision Term}

To further discuss the transport treatment we need an explicit form of the
collision term (\ref{Coll(kin)}), which is provided from the $\Phi$ functional
in the $-+$ matrix notation via the variation rules (\ref{varphi})
as
%
\def\tp{p'}\def\tm{m'}\def\tW{\widetilde{W}} 
\def\tR{\widetilde{R}}
\begin{eqnarray}
\label{Coll-var} 
C (X,p) =&& 
\frac{\delta\ii\Phi}{\delta\Ft(X,p)}\Ft(X,p)
-\frac{\delta\ii\Phi}{\delta\F(X,p)}\F(X,p) .
\end{eqnarray}
Here we assumed $\Phi$ be transformed into the Wigner representation and all
$\mp\ii\Gr^{-+}$ and $\ii\Gr^{+-}$ to be replaced by the Wigner-densities
$\Fd$ and $\Fdt$. Thus, the structure of the collision term can be inferred
from the structure of the diagrams contributing to the functional $\Phi$. To
this end, in close analogy to the consideration of ref.~\cite{KV}, we
discuss various decompositions of the $\Phi$-functional, from which the in-
and out-rates are derived. For the sake of physical transparency, 
we confine our treatment to the {\em local} case, where in the
Wigner representation all the Green's functions are taken at the 
same space-time coordinate $X$ and all non-localities, i.e. 
derivative corrections, are disregarded.  Derivative corrections 
give rise to memory effects in the collision term, which will be 
analyzed separately for the specific case of the triangle diagram.

Consider a given closed diagram of $\Phi$, at this level specified by a
certain number $n_{\lambda}$ of vertices and a certain contraction pattern.
This fixes the topology of such a contour diagram. It leads to $2^{n_\lambda}$
different diagrams in the $-+$ notation from the summation over all $-+$ signs
attached to each vertex.  Any $-+$ notation diagram 
of $\Phi$, which contains vertices of either sign, can
be decomposed into two pieces
in such a way that each of the
two sub-pieces contains vertices of only one type of sign\footnote{To
  construct the decomposition, just deform a given mixed-vertex diagram of
  $\Phi$ in such a way that all $+$ and $-$ vertices are placed left and
  respectively right from a vertical division line and then cut along this
  line.}
%
\begin{eqnarray}\label{decomp}\unitlength1mm
&&\ii \Phi_{\alpha\beta}=\unitlength1mm
\;
\def\ssp{\makebox(0,0)
    {\thinlines\put(-0.75,0){\line(1,0){1.5}}\put(0,-0.75){\line(0,0){1.5}}}}
\def\ssm{\makebox(0,0){\put(-0.75,0){\thinlines\line(1,0){1.5}}}}
\parbox{26mm}{\hspace*{-3mm}
\begin{fmfgraph*}(30,14)
\fmfforce{(0.5w,0.5h)}{v}
\fmfpen{thick}
\fmfstraight
\fmftopn{t}{11}
\fmfbottomn{b}{11}
\fmfleft{l}
\fmfright{r}
\fmf{plain,left}{b4,t4}
\fmf{plain,left}{t8,b8}
\fmf{plain}{b4,b5,lv1,lv2,lv3,lv4,lv5,lv6,t5,t4}
\fmf{plain}{b8,b7,rv1,rv2,rv3,rv4,rv5,rv6,t7,t8}
\fmffreeze
\fmf{fermion,label=$\ssp\hspace*{10mm}\ssm$,l.d=0}{lv1,rv1}
\fmf{fermion,label=$\ssp\hspace*{10mm}\ssm$,l.d=0}{lv2,rv2}
\fmf{fermion,label=$\ssp\hspace*{10mm}\ssm$,l.d=0}{lv3,rv3}
\fmf{fermion,label=$\ssp\hspace*{10mm}\ssm$,l.d=0}{rv4,lv4}
\fmf{fermion,label=$\ssp\hspace*{10mm}\ssm$,l.d=0}{rv5,lv5}
\fmf{fermion,label=$\ssp\hspace*{10mm}\ssm$,l.d=0}{rv6,lv6}
\fmf{phantom,label=\mbox{$\alpha\hspace*{14mm}\beta$},l.d=0}{l,r}
\end{fmfgraph*}}\;
    =\left(\alpha\left|
\Fd_1 ...  \Fdt'_{1} ...\right|\beta\right)
\\
    &\Rightarrow&
    \int\frac{\di^4 p_1}{(2\pi)^4}\cdots\frac{\di^4 p'_1}{(2\pi)^4}\cdots
    (2\pi)^4
    \delta^4\left(\sum_{i} p_i - \sum_{i} p'_i \right)
    V^*_{\alpha}
    \Fd_1 ... \Fdt'_{1} ...
    V_{\beta}
\nonumber
\end{eqnarray}
%
with $\Fd_1\cdots\Fd_m\Fdt'_1\cdots\Fdt'_{\tilde m}$ linking the two 
amplitudes.  The $V^*_{\alpha}(X;p_1,...p'_1,...)$ and 
$V_{\beta}(X;p_1,...p'_1,...)$ amplitudes represent multi-point vertex 
functions of only one sign for the vertices, i.e.  they are either entirely 
time ordered ($-$ vertices) or entirely anti-time ordered ($+$ vertices). 
Here we used the fact that adjoint expressions are complex conjugate to 
each other.  Each such vertex function is determined by normal Feynman 
diagram rules.  Applying the matrix variation rules (\ref{Coll-var}), we 
find that the considered $\Phi$ diagram gives the following contribution to 
the local part of the collision term (\ref{Coll(kin)}) 
%
\def\tp{p'}\def\tm{m'}\def\tW{\widetilde{W}} 
\def\tR{\widetilde{R}}
\begin{eqnarray}
\label{Coll-var-loc} 
\hspace*{-5mm}
&&C^{\scr{loc}} (X,p) \Rightarrow \frac{1}{2}
 \int \dpi{p_1} \cdots \dpi{\tp_1}\cdots 
R 
\left[\sum_{i} \delta^4(p_i-p)
-\sum_{i} \delta^4(\tp_i-p)\right]
\nonumber\\
&&\times 
\left\{
\Ft_1... \F'_1...
-
\F_1... \Ft'_1... 
\right\}
(2\pi)^4\delta^4\left(\sum_{i} p_i - \sum_{i} \tp_i \right).
\label{Multi-rate}
\end{eqnarray}
%
with the partial process rates
%
\begin{equation}\label{Rmm}
R(X;p_1,...p'_1,...)
=
\sum_{(\alpha\beta)\in \Phi} 
\Re\left\{
V^*_{\alpha}(X;p_1,...p'_1,...)
V_{\beta}(X;p_1,...p'_1,...)\right\}.
\end{equation}
%
The restriction to the real part arises, since with $(\alpha|\beta)$ also the
adjoint $(\beta|\alpha)$ diagram contributes to this collision term.
However these rates are not necessarily positive. In this
point, the generalized scheme differs from the conventional Boltzmann
kinetics. 

An important example of approximate $\Phi$ which we 
extensively use below is 
%
\begin{eqnarray}\unitlength1mm
\label{Phi-ring} 
\ii \Phi =\;\frac{1}{2}\;
\parbox{10mm}{
\begin{fmfgraph}(10,20)
\fmfpen{thick}
\fmfforce{(0.5w,0.5h)}{v}
\fmfforce{(0.5w,0.5h)}{v1}
\fmfstraight
\fmfright{r0,r1,r2,r3,r4}
\fmfleft{l0,l1,l2,l3,l4}
\fmf{phantom}{l1,v,r1}
\fmf{fermion,right=0.8,tension=0.45}{v,v}
\fmf{phantom}{l3,v1,r3}
\fmf{fermion,left=0.8,tension=0.45}{v1,v1}
\fmfdot{v1}
\end{fmfgraph}}
\;+\;\frac{1}{4}\;\;
\parbox{18mm}{
\begin{fmfgraph*}(18,18)
\fmfpen{thick}
\fmfsurroundn{v}{2}
\fmf{fermion,left=.25,tension=.7}{v1,v2,v1}
\fmf{fermion,right=.65,tension=.7}{v2,v1,v2}
\fmfdot{v1,v2}
\end{fmfgraph*}}
\;+\;\frac{1}{6}\;\;
\parbox{17mm}{
\begin{fmfgraph*}(18,18)
\fmfpen{thick}
\fmfsurroundn{v}{12}
\fmf{fermion,left=.25,tension=.7}{v4,v8,v12,v4}
\fmf{fermion,left=.25,tension=.7}{v12,v8,v4,v12}
\fmfdot{v4,v8,v12}
\end{fmfgraph*}}
\end{eqnarray}
%
where logarithmic factors due to the special features of the
$\Phi$-diagrammatic technique are written out explicitly,
cf. ref.~\cite{IKV99}. 
In this example we assume a system of fermions interacting via a two-body
potential $V=V_0 \delta(x-y)$, and, for the sake of simplicity, disregard its
spin structure. The $\Phi$ functional of Eq. (\ref{Phi-ring}) results in the 
following local collision term 
%
\unitlength1mm
\begin{eqnarray}
\label{C30} 
C^{\scr{loc}} 
&=& d^2
\int \dpi{p_1} \dpi{p_2} \dpi{p_3}
\left(
\left|\;\; 
\parbox{8mm}{
\begin{fmfgraph*}(8,8)
\fmfforce{(0.5w,0.5h)}{v}
\fmfpen{thick}
\fmfstraight
\fmftop{t1,t,t2}
\fmfbottom{b1,b2}
\fmf{fermion}{t1,v,b2}
\fmf{fermion}{b1,v,t2}
\fmfdot{v}
\fmflabel{$-$}{t}
\end{fmfgraph*}}
\;\;+ \;\;
\parbox{8mm}{
\begin{fmfgraph*}(8,15)\fmfkeep{twoint}
\fmfforce{(0.5w,0.5h)}{v}
\fmfpen{thick}
\fmfstraight
\fmftop{t1,t2,t3}
\fmfbottom{b1,b2,b3}
\fmf{fermion}{t1,t2,t3}
\fmf{fermion}{b1,b2,b3}
\fmf{fermion,left=.3,tension=.7}{t2,b2,t2}
\fmfdot{t2,b2}
\fmflabel{$-$}{t2}
\fmflabel{$-$}{b2}
\end{fmfgraph*}}
\;\;\right|^2 - 
\left|\;\;
\parbox{8mm}{
\fmfreuse{twoint}} \;\;\right|^2\right)
\nonumber 
\\[1mm]
&&\times 
\delta^4\left(p + p_1 - p_2 - p_3\right) 
\left(
\F_2\F_3 \Ft\Ft_1 -
\Ft_2\Ft_3 \F\F_1
\right),  
\end{eqnarray}
%
where $d$ is the spin (and maybe isospin) degeneracy factor. From this 
example one can see that the positive definiteness of transition rate is 
not evident. 

The first-order gradient corrections to the local collision term
(\ref{Multi-rate}) are called {\em memory} corrections. 
{\em Only diagrams of third and
higher order in the number of vertices give rise to memory effects}. 
In particular, only the last diagram of Eq. (\ref{Phi-ring}) gives rise to 
the memory correction, which is calculated in ref.~\cite{IKV99}.

\section{Entropy} \label{Entropy}

Compared to exact descriptions, which are time reversible, reduced description
schemes in terms of relevant degrees of freedom have access only to some
limited information and thus normally lead to irreversibility. 
In the Green's function formalism presented here the information loss
arises from the truncation of the exact Martin--Schwinger hierarchy, where the
exact one-particle Green's function couples to the two-particle Green's
functions, 
cf.  refs.~\cite{Kad62,BM}, which in turn are coupled to the three-particle
level, etc. This
truncation is achieved by the standard Wick decomposition, where all
observables are expressed through one-particle propagators and therefore
higher-order correlations are dropped. This step provides the Dyson's equation
and the corresponding loss of information is expected to lead to
a growth of entropy with time.  

We start with general manipulations which lead us to definition of the kinetic
entropy flow~\cite{IKV99}.  We multiply Eq. (\ref{keqk}) by $-\ln(\F/A)$, 
Eq. (\ref{keqkt}) by 
$(\mp)\ln(\Ft/A)$, take their sum, integrate it over $\di^4
p/(2\pi )^4$, and finally sum the result over internal degrees of freedom like
spin ($\Tr$). Then we arrive at the following relation
%
\begin{eqnarray}
\label{s-Eq.} 
\partial_\mu s^\mu_{\scr{loc}} (X) =\mbox{Tr}
\int \dpi{p} \ln \frac{\Ft}{\F} C (X,p),   
\end{eqnarray}
%
\begin{eqnarray}\label{entr-transp}
s^\mu_{\scr{loc}}  = \mbox{Tr}\int \dpi{p} 
A^{\mu}_{s} (X,p)\sigma (X,p),\,\,\, 
\end{eqnarray}
where 
\begin{eqnarray}\label{As}
\sigma (X,p) 
&=&\mp [1\mp f]\ln [1\mp f]-f\ln f ,
\\
A^{\mu}_{s} (X,p)&=&\frac{A \Gamma }{2}B^{\mu}
\end{eqnarray}
\begin{eqnarray}\label{B-j}
B^{\mu} = A \left[
\left(
2\pi^{\mu} - \frac{\partial \Re\Sa^R }{\partial p_\mu}
\right)
- M \Gamma^{-1}\frac{\partial \Gamma}{\partial p_\mu}
\right], 
\end{eqnarray}
cf. the corresponding drift term (proportional
to $\partial_{\mu}f$ in Eq. (\ref{keqk})). The zero-components of these
functions, $A^{0}_{s}$ and $B^{0}$, have a meaning of the entropy and flow
spectral functions, respectively, and satisfy the same sum rule as $A$.
If the considered particle is a resonance, like the $\Delta$ or
$\rho$-meson resonances in hadron physics, the $B^0$ function relates to the
energy variations of scattering phase shift of the scattering channel coupling
to the resonance in the discussed above virial limit. The value
 $s^0_{\scr{loc}}$ is interpreted as the local
(Markovian) part of the entropy flow. Indeed, the $s^0_{\scr{loc}}$ has 
proper thermodynamic and quasiparticle limits~\cite{IKV99}. However, to be 
sure that this is indeed the entropy flow we must prove the H-theorem for 
this quantity. 

First, let us consider the case, when memory corrections to the collision 
term are negligible. Then we can make use of the form
(\ref{Multi-rate}) of the local 
collision term. Thus, we arrive at the relation
%
\begin{eqnarray}
\label{s(coll)} 
&&\mbox{Tr}\int \dpi{p} \ln \frac{\Ft}{\F} C_{\scr{loc}} (X,p) 
\Rightarrow \mbox{Tr}
\frac{1}{2} 
 \int \dpi{p_1}\cdots \dpi{\tp_1}\cdots
R 
\nonumber 
\\
&&\times
\left\{\F_1... \Ft'_1 ... 
-
\Ft_1 ... \F'_1 ...\right\}
\ln\frac{\F_1 ... \Ft'_1...}
        {\Ft_1...\F'_1 ...}
(2\pi)^4
\delta^4\left(\sum_{i} p_i - \sum_{i} \tp_i \right).
\end{eqnarray}
%
In case all rates $R$ are
non-negative, i.e. $R\ge 0$, this expression is non-negative, since
$(x-y)\ln(x/y) \ge 0$ for any positive $x$ and $y$.  In particular,
$R\ge 0$ takes place for all $\Phi$-functionals up to two vertices.
Then the divergence of $s_{\scr{loc}}^\mu$ is non-negative 
%
\begin{eqnarray}\label{dmusmu>0}
\partial_\mu s^\mu_{\scr{loc}}(X)\ge 0,
\end{eqnarray}
%
which proves the $H$-theorem in this case with (\ref{entr-transp}) as the 
non-equilibrium entropy flow.
However, as has been mentioned above, we are unable to show that $R$
always takes non-negative values for all $\Phi$-functionals. 

If memory corrections are essential, the situation is even more involved. Let
us consider this situation again at the example of the $\Phi$ approximation 
given by Eq. (\ref{Phi-ring}). 
We assume that the fermion--fermion potential interaction is such that the
corresponding transition rate of the corresponding local 
collision term (\ref{C30}) is always non-negative, so that the $H$-theorem 
takes place in the local approximation, i.e. when we keep only 
$C^{\scr{loc}}$. 
Here we will schematically describe calculations of 
ref.~\cite{IKV99} which, to our opinion, 
illustrate a general strategy for the derivation of memory correction to the
entropy, provided the $H$-theorem holds for the local part.

Now Eq. (\ref{s-Eq.}) takes the form
%
\begin{equation}
\label{log-term}
\partial_\mu s^\mu_{\scr{loc}} (X) =\mbox{Tr}
\int \dpi{p} \ln \frac{\Ft}{\F} C^{\scr{loc}}
+\mbox{Tr}\int \dpi{p} \ln \frac{\Ft}{\F} C^{\scr{mem}},    
\end{equation}
%
where $ s^\mu_{\scr{loc}}$ is still the Markovian entropy flow defined by 
Eq.  (\ref{entr-transp}). Our aim here is to present the last term on the 
r.h.s. of Eq.  (\ref{log-term}) in the form of full $x$-derivative 
%
\begin{equation}
\label{ent-mar}
\mbox{Tr}\int \dpi{p} \ln \frac{\Ft}{\F} C^{\scr{mem}} 
=-\partial_{\mu} s_{\scr{mem}}^{\mu} (X) + \delta c_{\scr{mem}} (X)
\end{equation}
%
of some function $s_{\scr{mem}}^{\mu} (X)$, which we then interpret as a
non-Markovian correction to the entropy flow of Eq. (\ref{entr-transp}) 
plus a
correction ($\delta c_{\scr{mem}}$).
For the memory induced by the triangle diagram of Eq.(\ref{Phi-ring})
detailed calculations of ref.~\cite{IKV99} show that 
smallness of the $\delta 
c_{\scr{mem}}$, originating from small space--time gradients and small 
deviation from equilibrium, allows us to neglect this term as compared with 
the first term in r.h.s. of Eq. (\ref{ent-mar}).
Thus, we obtain
%
\begin{equation}
\label{H-der}
\partial_\mu \left(s^\mu_{\scr{loc}} + s_{\scr{mem}}^{\mu}\right) \simeq
\mbox{Tr}\int \dpi{p} \ln \frac{\Ft}{\F} C^{\scr{loc}} \geq 0, 
\end{equation}
%
which is the $H$-theorem for the non-Markovian kinetic equation under
consideration with $s^\mu_{\scr{loc}} + s_{\scr{mem}}^{\mu}$ as the proper
entropy flow. The r.h.s. of Eq. (\ref{H-der}) is non-negative 
provided the corresponding transition rate in the local 
collision term of Eq. (\ref{C30}) is non-negative.

The explicit form of $s_{\scr{mem}}^{\mu}$ is very complicated, see 
ref.~\cite{IKV99}. In equilibrium 
at low temperatures we get $s_{\scr{mem}}^0
\sim T^3 \ln T$ that gives the leading correction to the standard Fermi-liquid
entropy.  This is the famous correction~\cite{Baym91,Carneiro} 
to the specific heat of
liquid $^3$He. Since this correction is quite
comparable (numerically) to the leading term in the specific heat ($\sim T$),
one may claim that {\em liquid $^3$He is a liquid with quite strong memory
effects from the point of view of kinetics}.

\section{Pion-Condensate Phase Transition} 

As a further example for the role of finite width effects we consider the
phase transition dynamics into a pion-condensate.  The possible formation of
such a pion condensate in dense nuclear matter was initially suggested by A.B.
Migdal in his pioneering work \cite{Migdal1971}.  In realistic treatments of
this problem applied to equilibrated isospin-symmetric nuclear matter at low
temperatures $T\ll m_{\pi}$ the pion self-energy is determined by
nucleon--nucleon-hole, $\Delta$--nucleon-hole contributions corrected by
nucleon--nucleon correlations, $\pi\pi$ fluctuations and a residual
interaction \cite{MSTV}. A recent numerical analysis~\cite{Ph} within a
variational method with realistic two- and three-nucleon
interactions gave $\rho_c \simeq 2\rho_0$ for the critical density
of $\pi^+ ,\pi^- ,\pi^0$ condensation in symmetric nuclear matter and $\rho_c
\simeq 1.3\rho_0$ for $\pi^0$ condensation in neutron matter, with $\rho_0$
being nuclear saturation density.

In symmetric nuclear matter the pion condensate frequency vanishes
while the magnitude of condensate momentum $\vec{p}_c$, is approximately given
by the nucleon Fermi momentum $\mid\vec{p}_c \mid\simeq p_F$.  The critical
behavior of the system is determined by the effective pion gap
\begin{equation}\label{piongap}
\widetilde{\omega}^2 ({\vec p}_c)=\min_{\vec p}\left\{
m_\pi^2 +{\vec p}^2
+\mbox{Re}\Sigma^R_{\pi} (0, {\vec p},\varphi_\pi =0)\right\},
\end{equation}
\begin{figure}[h]
\centerline{
\includegraphics[height=7cm,clip=true]{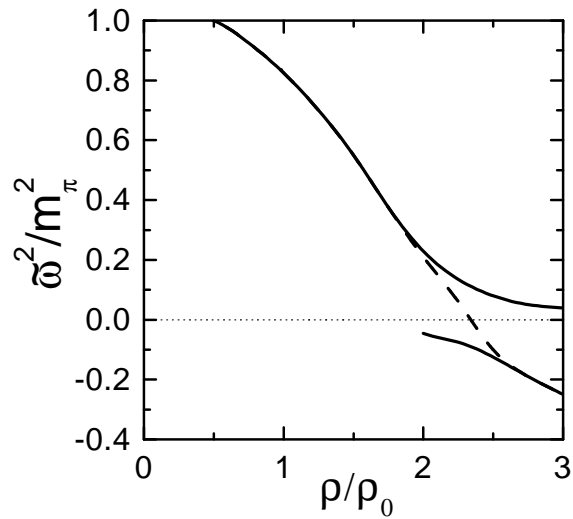}}
\caption{Effective pion gap (\ref{piongap}) 
versus nuclear density, see ref. \cite{MSTV}.}
\end{figure}
where the momentum $\vec{p}=\vec{p}_c$ corresponds to the minimum
of the gap at zero mean field $\varphi_\pi =0$ \cite{Migdl,MSTV}. 
Fig. 5 illustrates the behavior of the effective pion gap
$\widetilde{\omega}^2 ({\vec p}_c)$ as a function of the baryon density
$\rho$. At low densities, $\mbox{Re}\Sigma^R_{\pi}$ is small and one 
obviously has $\widetilde{\omega}^2 >0$.   
The dashed line in Fig. 5 describes the case where the $\pi\pi$ fluctuations
are artificially switched off and the phase transition turns out to be of
second order. At the critical point of the pion condensation ($\rho =\rho_c$)
this value of $\widetilde{\omega}^2$ with switched-off $\pi\pi$ fluctuations
changes its sign. 
In reality the $\pi\pi$ fluctuations are significant in the
vicinity of the critical point \cite{Dyugaev1975,VM81,Dyg1}. The corresponding
contribution to the pion self-energy behaves as 
$T/\widetilde{\omega}(\varphi_\pi,\vec{p}_c)$ at
$T>\mid\widetilde{\omega}^2(\varphi_\pi,\vec{p}_c)/m_\pi\mid$, and
$\widetilde{\omega}^2 ({\vec p}_c)$ does not cross the zero line at
all\footnote{Here we have used the $\widetilde{\omega}^2$ quantity that
already takes account of the pion mean field as explained below, 
cf. Eq. (\ref{m-f0}) and the definition of
$\widetilde{\omega}(\varphi_\pi,\vec{p}_c)$ after it.}.
Rather there are two branches (solid curves in Fig. 5) with positive and
respectively negative value for $\widetilde{\omega}^2 ({\vec p}_c)$ and the
transition becomes of the first order. Calculations of refs
\cite{Dyugaev1975,VM81,Dyg1} demonstrate that at 
$\rho >\rho_c$ the free energy of the state with $\widetilde{\omega}^2 ({\vec
  p}_0)>0$, where the pion mean field is zero, becomes larger than that of the
corresponding state with $\widetilde{\omega}^2 ({\vec p}_c)<0$ and a finite
mean field. Thus, at $\rho=\rho_c$ the first-order phase transition to the
inhomogeneous  pion-condensate state occurs. At 
$\rho >\rho_c$ the state with $\widetilde{\omega}^2 ({\vec p}_c)>0$ is
meta-stable and the state with $\widetilde{\omega}^2 ({\vec p}_c)<0$ and 
$\widetilde{\varphi}_\pi \neq 0$ becomes the ground state. 

Before we discuss a self-consistent scheme for a quantitative treatment 
of this
problem we like to qualitatively explain how the instability towards pion
condensation 
develops dynamically. To simplify the treatment we assume that the pion
density is low ($\rho_\pi\ll\rho$) and further use the fact that the pion is
much lighter than the nucleon ($m_{\pi}/m_N\simeq 1/7$). This allows us to
consider the pion sub-system as a light admixture in a heavy baryon
environment, neglecting the feedback of the pions onto the baryons. It
provides nucleon Green's functions unaffected by the pion distribution.
This very approximation was used in the first works exploring the possibility
of the pion condensation in dense nuclear matter
~\cite{Migdal1971,Migdal1972,Migdal1973}.  We will use it for
the pion retarded self-energy, thus neglecting the contribution from
pion fluctuations (see dashed curve in Fig. 5). Within the above
approximations the quantum kinetic equation (\ref{keqk}) for the pion
distribution $f_\pi$ in homogeneous and equilibrated baryon environment 
becomes 
\begin{equation}\label{bath}
\frac{1}{2} \Gm B_{\mu}\partial^{\mu}_X f_\pi =\Ge-\Gamma f_\pi.
\end{equation}
Here $B_{\mu}$ is defined in Eq. (\ref{B-j}) and all subscripts $\pi$ are
omitted, except for the pion distribution function $f_\pi$.

We like now to illustrate that the second branch in Fig. 5 with negative
$\widetilde{\omega}^2$ constructed under the assumption of vanishing mean
field is indeed unstable and becomes stabilized by a finite mean field.
The instability of the system can be discussed
considering a weak perturbation $\delta f_\pi$ of the pion distribution
$f^{(0)}_\pi=\left[\exp(p_0/T)-1\right]^{-1}$ which we assume equilibrated in
the rest frame of the system. Linearizing Eq. (\ref{bath}) we find
\begin{equation}\label{rel}
\frac{1}{2} B_{\mu}\partial^{\mu}_X \delta f_\pi 
+ \delta f_\pi =0, 
\end{equation}
with the solution
\begin{equation}\label{growth}
\delta f_\pi (t,p) = \delta f_0 (p) \exp\left[-2t/B_0(p)\right], 
\end{equation}
where for simplicity the initial fluctuation $\delta f_0 (p)$ of the pion
distribution is assumed to be space-independent.  Let us consider the case, 
where $p_0 \to 0$ and $\mid\vec{p}\mid\simeq p_F$. This four-momentum region,
being far from
the pion mass shell, is right the region, where the pion instability is
expected in symmetric nuclear matter. Here the real part of the pion
self-energy $\Re\Sigma^R$ is an even function of the pion energy $p_0$ while
the width is an odd function and proportional to $p_0$ for $p_0 \rightarrow
0$.  Using the results of refs. \cite{Migdl,MSTV} $2p_0 - \partial \Re\Sigma^R
/ \partial p_0 \rightarrow 0$, $\Gamma =\beta ({\vec p}) p_0$ and $\beta
({\vec p})\sim m_{\pi}$ for $p_0 \rightarrow 0$, we get $B_0 =\beta ({\vec
p})/\widetilde{\omega}^2 ({\vec p})$ from Eq.  (\ref{B-j}) and therefore
\begin{equation}\label{growth1}
\delta f_\pi (t,p_0=0,{\vec p}) = \delta f_0 (p_0=0,{\vec p})
\;\exp\left[-2\widetilde{\omega}^2({\vec p})t/\beta ({\vec p})\right].
\end{equation}
The above solution shows that for $\widetilde{\omega}^2 >0$ initial
fluctuations are damped, whereas they grow in the opposite case.  Thus,
the change of sign of $\widetilde{\omega}^2 (\vec{p}_c)$ leads to an
instability of the virtual pion distribution at low energies and momenta
$\simeq \vec{p}_c$.  The solution (\ref{growth1}) illustrates the important
role of the width in the quantum kinetic description. If the width would be
neglected in the quantum kinetic equation, one would fail to find the above
instability.

The growth of the pion distribution $\delta f_\pi$ is accompanied by a
growth of the condensate field $\varphi_\pi$.  Due to the latter the increase
of the virtual pion distribution slows down and finally stops when the mean
field reaches its stationary value.  Therefore, a consistent treatment of the
problem requires the solution of the coupled system of the quantum kinetic
equation (\ref{keqk}) and the mean field equation (\ref{fields}).  In order to
find the behavior of the virtual pion distribution one also has to include the
mean field contribution to the pion self-energy.  Considering only small mean
fields we keep terms of the lowest-order in $\varphi_\pi$.  Then
$\Sigma^R$ acquires an additional contribution $\Sigma^R
(\varphi_\pi)=\Sigma^R(\varphi_\pi=0)+\lambda_{\scr{eff}}
\mid\varphi_\pi\mid^2 $, where $\lambda_{\scr{eff}}$ denotes the total
in-medium pion-pion interaction. Within the same order the mean-field equation
becomes 
\begin{equation}\label{m-f0}
\left[\widetilde{\omega}^2(\vec{p}_c ) +\lambda_{\scr{eff}}
\widetilde{\varphi}_\pi^{2} (t) +\frac{1}{2}\beta (\vec{p}_c )\partial_t\right]
\widetilde{\varphi}_\pi (t)  =0. 
\end{equation}
Here we have assumed the simplest structure for the condensate field
$\varphi_\pi =\widetilde{\varphi}_\pi (t)\mbox{exp}(\ii \vec{p}_c\vec{r})$,
where $\widetilde{\varphi}_\pi (t)$ is a space-homogeneous real function which
varies slowly in time. Also one should do the replacement
$\widetilde{\omega}^2(\vec{p}_c)\rightarrow \widetilde{\omega}^2(\varphi_\pi,
\vec{p}_c)\equiv \widetilde{\omega}^2(\vec{p}_c)+ \lambda_{\scr{eff}}
\mid\varphi_\pi\mid^2$ in the above equations (\ref{bath}) - (\ref{growth1})
for the pion distribution.

The time dependence of $\widetilde{\varphi}$ can qualitatively be understood
inspecting the two limits of small and large times.
At short times the mean field is still small and one can neglect the
$\lambda_{\scr{eff}} \widetilde{\varphi}^{2}(t)$ term in
Eq. (\ref{m-f0}). Then the mean field
\begin{equation}\label{phi-grow}
\widetilde{\varphi}_\pi (t)=\widetilde{\varphi}_\pi (0)
\exp\left[-2\widetilde{\omega}^2 (\vec{p}_c)t/\beta (\vec{p}_c\right] 
\end{equation}
grows exponentially with time, just like the distribution function
(\ref{growth1}). Here $\widetilde{\varphi}_\pi (0)$ is an initial small
fluctuation of the field. 
At later times the solution of Eq. (\ref{m-f0}) approaches the stationary
limit  
$\widetilde{\varphi}_\pi \rightarrow \widetilde{\varphi}_\pi^{\scr{stat}}$
with  
\begin{equation}\label{phi-stat}
(\widetilde{\varphi}_\pi^{\scr{stat}})^2 =- \widetilde{\omega}^2(\vec{p}_c )
/\lambda_{\scr{eff}} (\vec{p}_c ).
\end{equation} 
Since simultaneously
$\widetilde{\omega}^2(\varphi_\pi, \vec{p}_c)= 
 \widetilde{\omega}^2(\vec{p}_c)+
\lambda_{\scr{eff}} \mid\varphi_\pi\mid^2 \rightarrow 0$, the change in the
pion distribution  $\delta f_\pi$ will saturate.
This stationary solution  $\widetilde{\varphi}_\pi^{\scr{stat}}$
is stable as can be seen from linearizing Eq. (\ref{m-f0}) around 
$\widetilde{\varphi}_\pi^{\scr{stat}}$
\begin{equation}
\widetilde{\varphi}_\pi (t)=\widetilde{\varphi}_\pi^{\scr{stat}}
-\delta\widetilde{\varphi}_0 
\exp\left[4\widetilde{\omega}^2(\vec{p}_c)t/\beta(\vec{p}_c)
\right],
\end{equation}
since the exponent is negative. Here $\delta\widetilde{\varphi}_0$ 
denotes an arbitrary initial space-homogeneous fluctuation.

The physics can again be cast into a $\Phi$-derivable form, where the
$\Phi$ functional should include at least the following diagrams
\unitlength=1mm
\begin{eqnarray}\label{phidiag}\nonumber
\Phi&=&\,\,
\parbox{10mm}{
\begin{fmfgraph*}(10,10)
\fmfpen{thick}
\fmfleft{l}
\fmfright{r}
\fmfforce{(0.0w,0.5h)}{l}
\fmfforce{(1.0w,0.5h)}{r}
\fmf{fermion,left=1.,tension=.5}{l,r}
\fmf{fermion,left=1.,tension=.5,label=\raisebox{-.6mm}{\hspace*{10mm}(a)}}{r,l}
\fmf{boson}{l,r}
\fmfdot{l,r}
\end{fmfgraph*}}
\,\, +\,\,
\parbox{20mm}{
\begin{fmfgraph*}(20,20)
\fmfpen{thick}
\fmfleft{l}
\fmfright{r}
\fmfforce{(0.0w,0.5h)}{l}
\fmfforce{(1.0w,0.5h)}{r}
\fmf{boson,left=.5,tension=.5}{l,r}
\fmfforce{(0.5w,0.1h)}{c1}
\fmf{fermion,left=.25,tension=.7}{l,c1}
\fmf{fermion,left=.25,tension=.7}{c1,l}
\fmf{fermion,left=.25,tension=.7}{c1,r}
\fmf{fermion,left=.25,tension=.7,label=\raisebox{-.7mm}{(b)}}{r,c1}
\fmfv{d.shape=square,d.fill=1,d.si=2thick}{c1}
\fmfdot{l,r}
\end{fmfgraph*}}
\,\,
+
\,\,
\parbox{30mm}{
\begin{fmfgraph*}(30,20)
\fmfpen{thick}
\fmfleft{l}
\fmfright{r}
\fmfforce{(0.0w,0.5h)}{l}
\fmfforce{(1.0w,0.5h)}{r}
\fmf{boson,left=.4,tension=.5}{l,r}
\fmfforce{(0.30w,0.1h)}{c1}
\fmfforce{(0.70w,0.1h)}{c2}
\fmf{fermion,left=.25,tension=.7}{l,c1}
\fmf{fermion,left=.25,tension=.7}{c1,l}
\fmf{fermion,left=.25,tension=.7}{c1,c2}
\fmf{fermion,left=.25,tension=.7}{c2,c1}
\fmf{fermion,left=.25,tension=.7}{c2,r}
\fmf{fermion,left=.25,tension=.7,label=\raisebox{-.7mm}{(c)}}{r,c2}
\fmfv{d.shape=square,d.fill=1,d.si=2thick}{c1,c2}
\fmfdot{l,r}
\end{fmfgraph*}}
\,\,\dots 
+\,\,
\parbox{30mm}{
\begin{fmfgraph*}(30,20)
\fmfpen{thick}
\fmfleft{l}
\fmfright{r}
\fmfforce{(0.0w,0.5h)}{l}
\fmfforce{(1.0w,0.5h)}{r}
\fmf{boson,left=.4,tension=.5}{l,r}
\fmfforce{(0.3w,0.1h)}{c1}
\fmfforce{(0.7w,0.1h)}{c2}
\fmf{fermion,left=.25,tension=.7}{l,c1}
\fmf{fermion,left=.25,tension=.7}{c1,l}
\fmf{fermion,left=.25,tension=.7}{c2,r}
\fmf{fermion,left=.25,tension=.7,label=\raisebox{-.7mm}{(d)}}{r,c2}
\fmfv{d.shape=square,d.fill=1,d.si=2thick}{c1,c2}
\fmfv{l={$\dots$},l.a=0,l.d=0.1w}{c1}
\fmfdot{l,r}
\end{fmfgraph*}}
\\ \nonumber\\[5mm] \label{phi-pion}
&+&\,\,
\parbox{11mm}{
\begin{fmfgraph}(10,20)
\fmfpen{thick}
\fmfforce{(0.5w,0.5h)}{v}
\fmfforce{(0.5w,0.5h)}{v1}
\fmfstraight
\fmfright{r0,r1,r2,r3,r4}
\fmfleft{l0,l1,l2,l3,l4}
\fmf{phantom}{l1,v,r1}
\fmf{fermion,right=0.8,tension=0.4}{v,v}
\fmf{phantom}{l3,v1,r3}
\fmf{fermion,left=0.8,tension=0.4}{v1,v1}
\fmfv{d.shape=square,d.fill=1,d.si=3thick}{v}
\end{fmfgraph}}
_{\mbox{(e)}}
 + \,\,
\parbox{11mm}{
\begin{fmfgraph}(10,20)
\fmfpen{thick}
\fmfforce{(0.5w,0.5h)}{v}
\fmfforce{(0.5w,0.5h)}{v1}
\fmfstraight
\fmfright{r0,r1,r2,r3,r4}
\fmfleft{l0,l1,l2,l3,l4}
\fmf{phantom}{l1,v,r1}
\fmf{boson,right=0.8,tension=0.4}{v,v}
\fmf{phantom}{l3,v1,r3}
\fmf{boson,left=0.8,tension=0.4}{v1,v1}
\fmfv{d.shape=diamond,d.fill=1,d.si=3thick}{v}
\end{fmfgraph}}
_{\mbox{(f)}}
 +\;\;
\parbox{17mm}{
\begin{fmfgraph}(15,15)
\fmfpen{thick}
\fmfforce{(0.5w,0.5h)}{l}
\fmfforce{(0.0w,0.5h)}{v}
\fmf{fermion,tension=.5}{l,l}
\fmf{boson}{l,v}
\fmfv{d.shape=tetragram,d.ang=45,d.fil=0,d.si=5thick}{v}
\fmfdot{l}
\end{fmfgraph}}
_{\raisebox{-3mm}{\mbox{(g)}}}
 +\,\,
\parbox{13mm}{
\begin{fmfgraph}(10,10)
\fmfpen{thick}
\fmfforce{(0.0w,0.0h)}{l}
\fmfforce{(1.0w,0.0h)}{r}
\fmfforce{(0.5w,0.5h)}{v}
\fmf{boson}{l,v}
\fmf{boson,tension=0.4}{v,v}
\fmf{boson}{v,r}
\fmfv{d.shape=tetragram,d.fil=0,d.si=5thick}{l,r}
\fmfv{d.sh=diamond,d.si=3thick}{v}
\end{fmfgraph}}
_{\raisebox{-5mm}{\mbox{(h)}}}
 +\,\,
\parbox{13mm}{
\begin{fmfgraph}(10,10)
\fmfpen{thick}
\fmfforce{(0.0w,1.0h)}{lu}
\fmfforce{(1.0w,1.0h)}{ru}
\fmfforce{(0.0w,0.0h)}{ld}
\fmfforce{(1.0w,0.0h)}{rd}
\fmfforce{(0.5w,0.5h)}{v}
\fmf{boson}{lu,v,ru}
\fmf{boson}{ld,v,rd}
\fmfv{d.shape=tetragram,d.fil=0,d.si=5thick}{lu,ru,ld,rd}
\fmfv{d.sh=diamond,d.si=3thick}{v}
\end{fmfgraph}}
_{\raisebox{-5mm}{\mbox{(i)}}}
\end{eqnarray}
Here bold and bold-wavy lines relate to baryon and pion Green's functions,
respectively, while the wavy line terminated by cross denotes the pion
condensate. Since in the broken phase the mean pion field mixes nucleon
with $\Delta$ configurations we adopt the $SU(4)$ formulation of the model,
introduced in ref. \cite{CDM}.  There one deals with a unified description of
baryons ($N$ and $\Delta$), based on $20\times 20$ matrix Hamiltonian in the
basis of $20$ $\Delta$--nucleon spin--isospin states. Thus, the solid lines
symbolize a unified propagator matrix for $\Delta$-resonance and nucleon. The
mixing is provided by condensate--baryon coupling (diagram (g)). Numerical
symmetry factors are omitted in Eq. (\ref{phidiag}).

Functional variation of $\Phi$ with respect to propagators provides the
corresponding self-energies. Diagrammatically this variation corresponds to
cutting and opening the respective propagator lines of the diagrams of $\Phi$
in Eq. (\ref{phidiag}). Thus diagrams (a) to (d), (f) and (h) 
contribute to the
pion self-energy. Diagram (a) accounts for the baryon particle--hole
contributions to the pion self-energy. It includes $NN^{-1}$, $\Delta N^{-1}$,
$N\Delta^{-1}$ and $\Delta\Delta^{-1}$ terms. The subsequent series of
diagrams (b) to (d) renormalizes baryon--pion vertex including baryon--baryon
correlations in terms of the Landau--Migdal parameter $g'$.  Diagram (f)
accounts for the pion fluctuations. It is proportional to $\sim
{T}/{\widetilde{\omega}(\varphi_\pi,\vec{p}_c)}$ and thus causes the
transition to be of 
first order. This becomes especially important for the case of heated and even
non-equilibrium dense matter, where the effective pion gap~\cite{VM81,Dyg1}
drops.  One should notice that pion fluctuation contributions are also present
in the particle-hole diagram (a) when opened perturbatively.  Diagram (h)
corresponds to pion interactions with the condensate which is responsible for
the stabilization of the condensate solution (\ref{phi-stat}).

Likewise cutting and opening the solid lines in $\Phi$ determines the baryon
self-energy which describes the feedback of the pions onto the baryonic
subsystem. This feedback is required for the conserving and thermodynamically
consistent treatment of the problem. Diagrams of the first line correspond to the modification of the
baryon motion by the multiple interaction with the pions corrected by
correlations. Diagram (e) generates a purely local interaction contribution
whereas diagram (g) with the coupling of the condensate to baryons leads to
the mixing of $N$ and $\Delta$. 

Variation of $\Phi$ with respect to the condensates (wavy line with cross)
determines the source term $J$ in the equation for the mean field
(\ref{fields}). The value $\lambda_{\scr{eff}}$ entering Eq. (\ref{m-f0}) is
generated by the the last two diagrams (h) and (i) of Eq. (\ref{phidiag}).

The kinetic description (\ref{keqk}) for the particle distribution  
together with the equation of motion for the mean field (\ref{fields}) 
is still insufficient for the numerical simulations
of the
dynamics of the phase transition. The reason is that 
the creation of seeds of the new phase, 
which initiate the growth of the mean field 
and the particle distribution, is due to
fluctuations, cf. Eqs. (\ref{growth1}) and (\ref{phi-grow}). However,
the scheme of Eqs. (\ref{keqk}) and 
(\ref{fields}) provides no sources of stochastic
fluctuations. Thus, it can only simulate the dynamics of one
of the phases rather than the transition between them.
The required stochastic
sources may be introduced into the transport theory in the spirit of the 
Boltzmann--Langevin approach developed in refs \cite{Ayik88,Rand90,Iv95} and
the stochastic interpretation of the Kadanoff--Baym equations \cite{Gre98}. 
The stochastic transport approach 
offers an appropriate 
framework for the description of the unstable dynamics by means of
a stochastic force in the mean field equation and
a stochastic collision term in 
the transport equation, 
which both act as a source for a continuous 
branching of the dynamical trajectories.

The above example shows that we really need the off-mass-shell kinetics to
describe the dynamics of the pion-condensate phase transition, since the
corresponding instability of the pion distribution function occurs far 
from the
pion mass shell, cf. Eq. (\ref{growth1}). Besides the conserving property and 
thermodynamic consistency of the $\Phi$-derivable approximation it also leads
us to the proper order of the phase transition.


\section{Summary and Prospects}

A number of problems arising in different dynamical systems, e.g. in
heavy-ion collisions, require an explicit
treatment of dynamical evolution of particles with finite mass-width. This was
demonstrated for the example of bremsstrahlung from a nuclear source, where
the soft part of the spectrum can be reproduced only provided the mass-widths
of nucleons in the source are taken explicitly into account. In this case the
mass-width arises due to collisional broadening of nucleons. Another examples
considered concern propagation of broad resonances 
(like $\rho$-meson and $\Delta$) in the
medium. Decays of $\rho$-mesons are an important source of di-leptons radiated
by excited nuclear matter. As shown, a consistent description of the
invariant-mass spectrum of radiated di-leptons can be only achieved if one
accounts for the in-medium modification of the $\rho$-meson width (more
precisely, its spectral function). The principle of {\em actio} = {\em
re-actio} was demonstrated on a pedagogical example 
when  there is only $\pi N\Delta$ coupling and  in 
the limit of a dilute nuclear matter.
We also expect a consistent
description of chiral $\sigma$-, $\pi$- condensates together with
fluctuations, as an immediate application of our results to multi-component
systems. 

We have argued that the Kadanoff--Baym equation within the 
first-order gradient approximation, slightly modified 
to make the set of Dyson's equations {\em exactly} consistent (rather than 
up to the second-order gradient terms), together with algebraic equation for
the spectral function provide a proper frame for a quantum
four-phase-space kinetic description that applies also to systems of unstable 
particles. 
The quantum four-momentum-space kinetic equation proves to be charge
and energy--momentum conserving and thermodynamically consistent, provided it 
is based on a $\Phi$-derivable approximation. The $\Phi$ functional also 
gives rise to a very natural representation of the collision term. 
Various self-consistent approximations are known since long time which do not
explicitly use the $\Phi$-derivable concept like self-consistent Born and
T-matrix approximations. The advantage the $\Phi$ functional method consists
in offering a regular way of constructing various self-consistent
approximations.

We have also addressed the question whether a closed
non-equilibrium system approaches the thermodynamic equilibrium during its
evolution. We obtained a definite expression for a local
(Markovian) entropy flow and were able to explicitly demonstrate the
$H$-theorem for some of the common choices of $\Phi$ approximations.  This
expression holds beyond the quasiparticle picture 
and thus generalizes the well-known Boltzmann kinetic
entropy.  Memory effects in the quantum four-momentum-space kinetics were
discussed and a general strategy to deduce   
memory corrections to the entropy was outlined.

At the example of pion-condensate phase transition in symmetric nuclear
matter we demonstrated important role of the width effects in the dynamics
and we formulated a self-consistent $\Phi$ derivable scheme for the transport
treatment of this problem. 
An interesting application of such self-consistent transport description
is possible to dynamics of the phase transition of a neutron star to the pion
or kaon condensate state accompanied by the corresponding neutrino burst. 
In view of the letter, another application concerns description of the neutrino
transport in supernovas and hot neutron stars during first few minutes of
their evolution. At an initial stage, neutrinos typically of thermal energy,
produced outside (in the mantel) and inside the neutron-star core, are trapped
within the regions of production.  However, coherent effects in neutrino
production and their rescattering on nucleons \cite{KV} reduce the
opacity of the nuclear-medium and may allow for soft neutrinos to escape the
core and contribute to the heating off the mantle. The extra energy transport
may be sufficient to blow off the supernova's mantle in the framework of
the shock-reheating mechanism \cite{Wilson}. The description of the neutrino
transport in the semi-transparent region should therefore be treated with the
due account of mass-widths effects.

Further applications concern relativistic plasmas, such as QCD and QED
plasmas. The plasma of deconfined quarks and gluons was present in the early
Universe, it may exist in cores of massive neutron stars, and may also be
produced in laboratory in ultra-relativistic nucleus--nucleus collisions.  All
these systems need a proper treatment of particle transport.  Perturbative
description of soft-quanta propagation suffers of infrared divergences and one
needs a systematic study of the particle mass-width effects in order to treat
them, cf. ref. \cite{KV}.  A thermodynamic $\Phi$-derivable approximation
for hot relativistic QED plasmas---a gas of electrons and positrons in a
thermal bath of photons---was recently considered in ref.~
\cite{Vanderheyden}. Their treatment may be also applied to the
high-temperature super-conductors and the fractional quantum Hall effect
\cite{Halp93,Ioff89}. Formulated above 
approach allows for a natural generalization of such
a $\Phi$-derivable schemes to the dynamical case.
\\


\noindent
{\bf Acknowledgment:} We are grateful to G. Baym, G.E. Brown, P. Danielewicz,
H. Feldmeier, B.  Friman, C. Greiner, E.E. Kolomeitsev, P.C. Martin, U. Mosel
and S. Leupold for fruitful discussions. Two of us (Y.B.I. and D.N.V.) highly
appreciate the hospitality and support rendered to us at Gesellschaft f\"ur
Schwerionenforschung.  This work has been supported in part by DFG (project
436 Rus 113/558/0).  Y.B.I and D.N.V. were partially supported by RFBR grant
NNIO-00-02-04012.  Y.B.I. was also partially supported by RFBR grant
00-15-96590.

\end{fmffile}

\begin{thebibliography}{99}
\bibitem{Landau}
L.D. Landau, {\it ZhETF} {\bf 30},  1058 (1956) [
{\it Sov. Phys. JETP} {\bf 3},  920 (1956)]. 
\bibitem{Mqp}
A.B. Migdal, {\it Nuclear Theory: the Quasiparticle Method},
W.A. Benjamin, N.Y. 1968.
\bibitem{M} 
A.B. Migdal, {\it Theory of Finite Fermi Systems and properties of Atomic 
Nuclei},
Wiley and Sons, New York 1967 
(Russ. ed. 1965), 2-nd edition, Moscow, Nauka 1983 in
Russian.
\bibitem{LP1981} 
E.M. Lifshiz and L.P. Pitaevskii, {\it
Statistical Physics P2, V9, Pergamon 1980}.
\bibitem{Migdal1971}
A.B. Migdal, {\it ZhETF } {\bf 61}, 2209  (1971) 
[{\it Sov. Phys. JETP } {\bf 34},  1184 (1972)].
\bibitem{Migdal1972}
A.B. Migdal, {\it ZhETF} {\bf 63},  1993 (1972)
[{\it Sov. Phys. JETP} {\bf 36},  1052 (1973)].
\bibitem{Migdal1973}
A.B. Migdal, {\it Phys. Rev. Lett.}
{\bf 31},  257 (1973); {\it Nucl. Phys.}
{\bf A210},  421 (1973); {\it  Phys. Lett.}
{\bf 47B},  96 (1973), {\bf 52B},  172, 264 (1974).
\bibitem{Migdl}
A.B. Migdal, {\it Rev. Mod. Phys.} {\bf 50},  107 (1978).
\bibitem{MSTV}
A. B. Migdal, E. E. Saperstein, M. A. Troitsky and D.N.Voskresensky,
{\it Phys. Rep.}  {\bf 192},  179 (1990).
\bibitem{Vos93}
D.N. Voskresensky, {\it Nucl. Phys.} {\bf A555},  293 (1993).
\bibitem{LandauP} L. D. Landau and I. Pomeranchuk, {\it
Dokl. Akad. Nauk SSSR} {\bf 92},  553, 735 (1953); also in ''Collected
Papers of Landau'', ed. Ter Haar (Gordon \& Breach, 1965) papers 75 -
77.
\bibitem{Migdal} A. B. Migdal, {\it Phys. Rev.} {\bf 103},
1811 (1956) [ {\it Sov. Phys. JETP} {\bf 5},  527 (1957)].
\bibitem{eSLAC} P. L. Anthony, et al., {\it Phys. Rev. Lett.} {\bf 75},
1949 (1995); S. Klein, 
{\it Rev. Mod. Phys.} {\bf 71},  1501 (1999). 
\bibitem{Schw}
J. Schwinger, {\em J. Math. Phys.} {\bf 2}, 407 (1961).
\bibitem{Kad62}
L.P. Kadanoff and G. Baym, {\it Quantum Statistical
Mechanics} (Benjamin, 1962).
\bibitem{Keld64}
L.P. Keldysh, {\em ZhETF} {\bf 47}, 1515 (1964) [{\em Sov. Phys. JETP} {\bf
20},  1018 (1965)].
\bibitem{KadB} G. Baym and L.P. Kadanoff, {\it Phys. Rev. } {\bf 124},  287
(1961).
\bibitem{Baym}
G. Baym, {\em Phys. Rev.} {\bf 127}, 1391 (1962).
\bibitem{Luttinger}
J. M. Luttinger and J. C. Ward, {\it Phys. Rev. } {\bf 118},  1417 (1960).
\bibitem{Abrikos}
A.A. Abrikosov, L.P. Gorkov, I.E. Dzyaloshinski, 
{\it Methods of Quantum Field Theory in Statistical Physics },
Dover Pub., INC. N.Y., 1975.
\bibitem{IKV}
Yu.B. Ivanov, J. Knoll, and D.N. Voskresensky,  
{\em Nucl. Phys.} A {\bf 657}, 413 (1999).
\bibitem{IKV99}
Yu.B. Ivanov, J. Knoll and D.N. Voskresensky, 
{\em Nucl. Phys.} A {\bf 672}, 313 (2000).
\bibitem{KV} J. Knoll and D.N. Voskresensky, {\em Ann. Phys.} {\bf 249},
532 (1996); {\em Phys. Lett.} B {\bf 351}, 43 (1995).
\bibitem{Hees}
H. van Hees and J. Knoll, proceedings of the Intern. Workshop XXVIII on Gross
Properties of Nuclei and Nuclear Excitations, Hirschegg, Austria, Jan
16-22,2000, hep-ph/0002087, and to be published. 
\bibitem{Weinhold} W. Weinhold,
B.L. Friman, and W. N\"orenberg, {\em Acta Phys. Pol.} {\bf 27}, 3249 (1996);
{\em Phys. Lett.} B {\bf 433}, 236 (1998).
\bibitem{DGK} {J. Knoll and C. Guet}, {\em Nucl. Phys.} A {\bf 494},
 334 (1989); {M. Durand and J. Knoll}, {\em Nucl. Phys.} A {\bf 496}, 539
(1989).
\bibitem{VS87}
D. N. Voskresensky and A. V. Senatorov, {\it Yad. Fiz.} {\bf 45},  657 (1987) 
[in Engl. translation {\it Sov.J.Nucl.Phys. }{\bf 45}, 411 (1987).
\bibitem{Knoll-Erice}
J. Knoll, proceedings of the Erice School on Nuclear
Physics in Erice, Sicily, Italy, September 17 -25 1998, published in Progress
in Particle and Nuclear Physics {\bf 42}, 177 (1999).
\bibitem{HFN}M. Herrmann, B.L. Friman, and W. N\"orenberg,
{\em Nucl. Phys.} A {\bf 560}, 411 (1993).
\bibitem{Rapp}R. Rapp, G. Chanfray, and J. Wambach, {\em Nucl. Phys.} A {\bf
    617}, 472 (1997). 
\bibitem{Mosel}S. Leupold, U. Mosel, {\em Phys. Rev.} C {\bf 58}, 2939 (1998). 
\bibitem{Klingl}F. Klingl, N. Kaiser, and W. Weise,  {\em Nucl. Phys.} A {\bf
    624}, 527 (1997).
\bibitem{FrimanPirner}B.L. Friman and H.-J. Pirner, {\em Nucl. Phys.} A {\bf
    617}, 496 (1997)
\bibitem{FLW} B. Friman, M. Lutz and G. Wolf, nucl-th/9811040, in Proc.
"Baryons 98", Bonn, 22.-26. September 1998, World Scientific 1999, 663-674 and
in Proc. "Intern. Workshop XXVII on Gross Properties of Nuclei and
Nuclear Excitations", Hirschegg, January 2000; nucl-th/0003012.
\bibitem{DB} P. Danielewicz and G.F. Bertsch, {\em Nucl. Phys.} A  {\bf
533}, 712 (1991).
\bibitem{Effenberger99} M. Effenberger, E.L. Bratkovskaya and U. Mosel, {\em
Phys. Rev. } C {\bf 60}, 051901 (1999).
\bibitem{Effenberger99-1}M. Effenberger, E.L. Bratkovskaya, W. Cassing,
U. Mosel, {\em Phys. Rev. } C {\bf 60} 027601 (1999).
\bibitem{BethU} E. Beth, G.E. Uhlenbeck, {\it Physica} 4,  915 (1937).
\bibitem{Huang} K. Huang, "Statistical Mechanics", Wiley, New York (1963).
\bibitem{DMB} R. Dashen, S. Ma, H.J. Bernstein, {\it Phys.Rev.} 187,  345 
(1969).
\bibitem{Mekjian} A. Z. Mekjian, {\it Phys. Rev.} C {\bf 17},  1051 (1978).
\bibitem{VPrakash}R. Venugopalan and M. Prakash, {\it Nucl. Phys.} 
{\bf A 456}, 718 (1992).
\bibitem{Lenz} W. Lenz, {\it Z. Phys.} {\bf 56},  778 (1929). 
\bibitem{Land77}
L.D. Landau and E.M. Lifshiz, {\it Quantum Mechanics}, Part 3,
Pergamon press, 1977.
\bibitem{Dover71}C.B. Dover, J. H\"ufner and R.H. Lemmer, {\em
Ann. Phys. (NY)} {\bf 66}, 248 (1971).
\bibitem{Lutz94} M. Lutz, A. Steiner and W. Weise, {\em Nucl. Phys. A} {\bf
574}, 755 (1994).
\bibitem{DP}P. Danielewicz and S. Pratt, {\it Phys.Rev.}  {\bf C53},  249
(1996).
\bibitem{BM}
W. Botermans and R. Malfliet, {\em Phys. Rep.} {\bf 198}, 115 (1990).
\bibitem{Leupold00}S. Leupold, {\em Nucl. Phys. A} {\bf 672}, 475 (2000).
\bibitem{Juchem00}W. Cassing and S. Juchem,  {\em Nucl.Phys.A} {\bf 672}, 417
(2000). 
\bibitem{Bez}
B. Bezzerides and D.F. DuBois, {\em Ann. Phys. (N.Y.)} {\bf 70}, 10 (1972).
\bibitem{LSV}
P. Lipavsky, V. Spicka and B. Velicky, {\em Phys. Rev.} D {\bf 34},
6933 (1986).
\bibitem{LipS}  
V. Spicka and P. Lipavsky, {\em Phys. Rev. Lett.}
{\bf 73}, 3439 (1994);  {\em Phys. Rev.} B {\bf 52}, 14615 (1995).
\bibitem{Kraft}
W.D. Kraeft, D. Kremp, W. Ebeling and G. R\"{o}pke, {\it Quantum
Statistics of Charged Particle Systems } (Akademie-Verlag, Berlin, 1986).
\bibitem{Bonitz} 
M. Bonitz, {\it Quantum Kinetic Theory} (Teubner, Stuttgart/Leipzig, 1998).
\bibitem{Bornath}
Th. Bornath, D. Kremp, W. D. Kraeft, and M. Schlanges, {\em Phys. Rev.} E
{\bf 54}, 3274 (1996). 
\bibitem{SCFNW}
M. Sch\"{o}nhofen, M. Cubero, B. Friman, W. N\"{o}renberg and Gy. Wolf,
{\em Nucl. Phys.} A {\bf 572}, 112 (1994).
\bibitem{Jeon}
S. Jeon and L.G. Yaffe, {\em Phys. Rev.} D {\bf 53}, 5799 (1996). 
\bibitem{VBRS}
D.N. Voskresensky, D. Blaschke, G. R\"{o}pke and H. Schulz,
{\em Int. Mod. Phys. J.} E {\bf 4}, 1 (1995).
\bibitem{Cornwall} J.M. Cornwall, R. Jackiw and E. Tomboulis,
{\em Phys. Rev.} D {\bf 10}, 2428 (1974).
\bibitem{Baym91}
G. Baym and C. Pethick, {\it Landau Fermi--Liquid Theory} (John
 Wiley and Sons, INC, N.Y., 1991).
\bibitem{Carneiro}
G.M. Carneiro and C. J. Pethick, {\em Phys. Rev.} B {\bf 11}, 1106 (1975).
\bibitem{Ph}  Akmal, A., Pandharipande, V. R., and 
Ravenhall, D. G., {\it Phys.~Rev.}{\bf C 58}, 1804 (1998).
\bibitem{Dyugaev1975}A.M. Dyugaev,{\it  Pisma v ZhETF} {\bf 22},  181 (1975).
\bibitem{VM81}
D.N. Voskresensky and I.N. Mishustin, {\it Pisma v ZhETF} {\bf 34},  
317 ( 1981) [in Engl.: {\it JETP Lett.} {\bf 34},  303 (1981)];
{\it Yad. Fiz.} {\bf 35}, ( 1982) [in Engl.:{\it Sov. J. Nucl. Phys.}
{\bf 35},  667 (1982)].
\bibitem{Dyg1}
A.M. Dyugaev, {\it Pisma v ZhETF.} {\bf 35},  341 (1982);
{\it ZhETF} {\bf 83},  1005 (1982); {\it Yad. Fiz.} {\bf 38},  1131 (1983)
[in Engl.:{\it Sov. J. Nucl. Phys.}
{\bf 38},  680 (1983)].
\bibitem{CDM}
D. Campbell, R. Dashen and J. Manassash,
Phys. Rev. {\bf D12},  979, 1010 (1975).
\bibitem{Ayik88}
        S.~Ayik and C.~Gregoire,
        Phys. Lett. {\bf B212},  269 (1988);
        Nucl. Phys. {\bf A513},  187 (1990). 
\bibitem{Rand90} 
        J.~Randrup and B.~Remaud,
        Nucl. Phys. {\bf A514},  339 (1990). 
\bibitem{Iv95}
Yu.B. Ivanov and S. Ayik, Nucl.Phys. {\bf A593},   233 (1995). 
\bibitem{Gre98}
C. Greiner and S. Leupold, Ann. Phys. {\bf 270},  328 (1998). 
\bibitem{Wilson}
H.A. Bethe and J.R. Wilson, Astroph. J. {\bf 295},  14 (1985).
\bibitem{Vanderheyden} B. Vanderheyden and G. Baym, J.
  Stat. Phys. {\bf 93},  843 (1998).
\bibitem{Halp93}
B.I. Halperin, P.A. Lee and N. Read, Phys Rev. {\bf B47},  7312 (1993).
\bibitem{Ioff89}
L.B. Ioffe and A.I. Larkin, {\bf B39},  8988 (1989). 
\end{thebibliography}
\end{document}